\documentclass[longbibliography,noeprint,noshowpacs,nopreprintnumbers,twocolumn,pra,showpacs,superscriptaddress,amsmath,amssymb,citeautoscript,aps,10pt]{revtex4-2}
\usepackage{graphicx}
\usepackage{dcolumn}
\usepackage{color}
\usepackage{bm}
\usepackage[hidelinks]{hyperref}
\hypersetup{
    colorlinks,
    citecolor=blue,
    filecolor=blue,
    linkcolor=blue,
    urlcolor=blue
}

\graphicspath{{Figures/}{./Figures/}{.}}

\DeclareMathOperator{\tr}{tr}
\DeclareMathOperator{\C}{\mathcal{C}}
\DeclareMathOperator{\hc}{\mathcal{H}}
\DeclareMathOperator{\hh}{\hat{H}}
\DeclareMathOperator{\cs}{\mathrm{cs}}
\renewcommand{\vec}[1]{\boldsymbol #1}

\newcommand{\im}{\text{i}}

\begin{document}

\title{Fisher zeroes and dynamical quantum phase transitions\\ for two- and three-dimensional models}

\author{Tomasz Mas{\l}owski}
\affiliation{The Faculty of Mathematics and Applied Physics, Rzesz\'ow University of Technology, al.~Powsta\'nc\'ow Warszawy 6, 35-959 Rzesz\'ow, Poland}
\author{Hadi Cheraghi}
\affiliation{Computational Physics Laboratory, Physics Unit,
Faculty of Engineering and Natural Sciences, Tampere University, Tampere FI-33014, Finland}
\affiliation{Helsinki Institute of Physics, University of Helsinki FI-00014, Finland}
\author{Jesko Sirker}
\affiliation{Department of Physics and Astronomy and Manitoba Quantum Institute, University of Manitoba, Winnipeg, Canada R3T 2N2}
\author{Nicholas Sedlmayr}
\email[e-mail:]{sedlmayr@umcs.pl}
\affiliation{Institute of Physics, M. Curie-Sk\l{}odowska University, 20-031 Lublin, Poland}

\date{\today}

\begin{abstract}
Dynamical quantum phase transitions are non-analyticities in a dynamical free energy (or return rate) which occur at critical times. Although extensively studied in one dimension, the exact nature of the non-analyticity in two and three dimensions has not yet been fully investigated. In two dimensions, results so far are known only for relatively simple two-band models. Here we study the general two- and three-dimensional cases. We establish the relation between the non-analyticities in different dimensions, and the functional form of the densities of Fisher zeroes. We show, in particular, that entering a critical region where the density of Fisher zeroes is non-zero at the boundary always leads to a cusp in the derivative of the return rate while the return rate itself is smooth. We illustrate our results by obtaining analytical results for exemplary two- and three-dimensional models.  
\end{abstract}

\maketitle


\section{Introduction}\label{sec:intro}

In one-dimensional systems a time analogue of a quantum phase transition called a dynamical quantum phase transition (DQPT), at which a suitably defined free energy becomes non-analytic, have been extensively studied~\cite{Heyl2013,Andraschko2014,Heyl2018a,Sedlmayr2019a}. In this case the ``free energy'' is the return rate, which is to the Loschmidt amplitude as the free energy is to the partition function. For the purposes here the Loschmidt amplitude can be understood as the overlap between an initial and a time evolved state. The non-analyticity can be traced to zeroes of the Loschmidt amplitude when the time evolved state becomes orthogonal to the initial state. Furthermore, by extending to complex time, these zeroes can be understood as those times when Fisher zeroes, i.e.,~the zeroes of the Loschmidt amplitude with a complex argument, cross the real time axis. 

Following the introduction of DQPTs, much theoretical work followed~\cite{Karrasch2013,Heyl2014,Heyl2015,Sharma2015,Halimeh2017,Homrighausen2017,Halimeh2018,Shpielberg2018,Zunkovic2018,Srivastav2019,Huang2019,Gurarie2019,Abdi2019,Puebla2020,Link2020,Sun2020,Rylands2021a,Trapin2021,Yu2021,Halimeh2021,Halimeh2021a,DeNicola2021,Cheraghi2021,Cao2021,Bandyopadhyay2021,Cheraghi2023,Wong2023b} and experiments in ion traps, cold atomic gases, and quantum simulator platforms have been performed~\cite{Jurcevic2017,Flaschner2018,Zhang2017b,Guo2019,Smale2019,Nie2020,Tian2020}. DQPTs are of interest as a relatively simple example of non-equilibrium physics which can be studied systematically, and they can give information about systems beyond their equilibrium phase diagrams~\cite{Vajna2014,Andraschko2014,Karrasch2017,Jafari2017,Cheraghi2018,Jafari2019,Wrzesniewski2022}. The idea has also been extended to Floquet systems~\cite{Sharma2014,Yang2019,Zamani2020,Zhou2021a,Jafari2021,Hamazaki2021,Zamani2022,Luan2022,Jafari2022} and to mixed states, finite temperatures, and open or dissipative systems~\cite{Mera2017,Sedlmayr2018b,Bhattacharya2017a,Heyl2017,Abeling2016,Lang2018,Lang2018a,Kyaw2020,Starchl2022,Naji2022,Kawabata2023}. A particular focus has been on DQPTs in topological matter~\cite{Schmitt2015,Vajna2015,Jafari2016,Bhattacharya2017,Jafari2017a,Sedlmayr2018,Jafari2018,Zache2019,Maslowski2020,Okugawa2021,Rossi2022,Maslowski2023,Maslowski2024} where dynamical order parameters can be defined~\cite{Budich2016,Heyl2017,Bhattacharya2017a} and a dynamical bulk-boundary correspondence has also been introduced~\cite{Sedlmayr2018,Sedlmayr2019a,Maslowski2020,Maslowski2023,Maslowski2024}. Most work has concentrated on one dimension and for topological matter on simple two-band models. However multi-band models~\cite{Huang2016,Jafari2019,Mendl2019,Maslowski2020}, and two-dimensional systems~\cite{Schmitt2015,Vajna2015,Bhattacharya2017,DeNicola2022,Hashizume2022,Brange2022} have also been studied, and recently DQPTs in a three-dimensional model have been considered as well~\cite{Kosior2024}.

Fisher zeroes occur in the complex time plane when the Loschmidt amplitude is zero. For a bulk system they can be parameterised by the crystal momentum: for each point in the Brillouin zone there are a set of Fisher zeroes. In one dimension we therefore have a set of lines, and if they pass the real time axis they cause DQPTs. These DQPTs can be related therefore to critical momenta. In two dimensions the Fisher zeroes form planes, and there are now extended regions labelled by sets of critical momentum loops~\cite{Schmitt2015,Vajna2015}. The density of the zeroes if such a region crosses the real time axis determines the nature of the non-analyticity that occurs for the return rate or its derivatives. Understanding this in detail and generality is one of the main aims of this article. For a simple two-band model it is understood that cusps occur in the first derivative of the return rate at the beginning and end of the critical region~\cite{Schmitt2015,Vajna2015}. The distribution of Fisher zeroes in two-dimensional DQPTs has been studied numerically~\cite{Sacramento2024}, though again the focus is on two-band models, for which the behaviour may not be generic. In this paper we extend these ideas to general two-dimensional and three-dimensional models. We give a general argument for the relation between the behaviour of the Fisher zeroes crossing the real time axis, and the nature of the non-analyticity of the return rate, for any dimension. For two-band models we reiterate the argument that the density of Fisher zeroes diverges at the boundaries of the critical region, offering a new derivation, and relate this to the behaviour of the return rate. We illustrate our findings with results for two exemplary models where analytical formulae can be derived.

This paper is organised as follows. In Sec.~\ref{DQPTs} we define DQPTs and Fisher zeroes, relating the behaviour of the two in one, two, and three dimensions. We give further details for the two-dimensional case where explicit calculations of the density of Fisher zeroes are possible, and derive results for generic two-band models. In Sec.~\ref{sec:results}, we illustrate our general results by considering specific two- and three-dimensional topological models. We derive their equilibrium topological phase diagrams, and results for the Fisher zeroes, critical momenta, and DQPTs. In Sec.~\ref{sec:con} we conclude. Further details on the topological phase diagrams, some explicit analytical formulae, and additional examples of DQPTs can be found in the appendices.

\section{Fisher zeroes and dynamical quantum phase transitions}
\label{DQPTs}

Here we want to discuss the general setup and the differences between quantum quenches in one, two, and three dimensions. We think about a system prepared in the ground state $|\Psi_0\rangle$ of a Hamiltonian $\hh_0$. This state is then time evolved by a Hamiltonian $\hh_1$ which is typically obtained from $\hh_0$ by a sudden change of parameters, a so-called quantum quench. We are interested in the overlap of the time evolved with the original state
\begin{equation}
\label{Loschmidt}
L(t)=\langle\Psi_0|e^{-i\hh_1t}|\Psi_0\rangle
\end{equation}
which is also called the Loschmidt amplitude. The Loschmidt echo, measured for example in nuclear magnetic resonance, is obtained as $\mathcal{L}(t)=|L(t)|^2$. With increasing system size $N$, we will in general have an orthogonality catastrophe. I.e., the time evolved state $|\Psi_0(t)\rangle=e^{-i\hh_1t}|\Psi_0\rangle$ will have an overlap with $|\Psi_0\rangle$ which is exponentially small in $N$. This is the motivation to define the return rate as
\begin{equation}
    \label{return}
    l(t)=-\lim_{N\to\infty}\frac{1}{2N}\ln\mathcal{L}(t)
    = \frac{1}{2}\lim_{N\to\infty}\left[f(t)+f^*(t)\right] 
\end{equation}
with $f(t)=-\frac{1}{N}\ln L(t)$. We can consider $L(t)$ as being a function of the complex parameter $z=it$. For a finite system, $L(z)$ is an entire function in the complex plane. This can be seen by inserting eigenfunctions of $\hh_1$ in Eq.~\eqref{Loschmidt}. The Weierstrass factorization theorem then allows us to write \cite{Heyl2013}
\begin{align}
\label{fz}
L(z)=&e^{g(z)} \prod_j \left(1-\frac{z}{z_j}\right)\textrm{ and} \\ 
f(z)=&-\frac{g(z)}{N}\underbrace{-\frac{1}{N}\sum_j\ln\left(1-\frac{z}{z_j}\right)}_{\tilde f(z)}\,,  \nonumber
\end{align}
where $g(z)$ is an entire function and $\tilde f(z)$ denotes the part of $f(z)$ which contains the Fisher zeroes $z_j$ with $L(z_j)=0$. There are now important differences about the localization of the Fisher zeroes in the complex plane between the one-dimensional, and the two-dimensional and three-dimensional cases which we will discuss in the following. 

A note on nomenclature: we will refer to the extended times where Fisher zeroes cross the real time axis as \emph{critical regions}. By \emph{critical times} we mean the times at which there is a non-analyticity in the return rate.

\subsection{One dimension}
Let us first discuss the one-dimensional case. For periodic boundary conditions, we can express the Hamiltonian $\hh_1$ in  Eq.~\eqref{Loschmidt} as a function of momentum $k$. In the thermodynamic limit, the Fisher zeroes $z_p(k): \mathbb{R}\to\mathbb{C}$ which we parametrize as $z_p(k)=a_p(k)+i b_p(k)$ with real functions $a_p(k),\, b_p(k)$ then form curves in the complex plane labelled by an integer $p$. The return rate $l(t)$ will show a singularity whenever these curves cross the real time axis, i.e., if $a_p(k)=0$. Near such a critical time $t_c$, we can parametrize the curve of Fisher zeroes as $z(k)=it_c\pm e^{i\alpha}\tau(k)$ with $\tau(k)$ real which will cross the imaginary axis at an angle $\alpha$. To keep the discussion general, we can, furthermore assume that close to the critical time the density of Fisher zeroes follows a power law $\tau(k)\sim k^n$. Turning the sum into an integral in Eq.~\eqref{fz} we obtain
\begin{align}
    \label{fz2}
    \tilde f(t) =& -\int \frac{dk}{2\pi}\ln\left(1-\frac{it}{it_c\pm e^{i\alpha}k^n} \right) \\
    =& -\frac{1}{2\pi}\int dy \underbrace{\left| \frac{dy}{dk}\right|^{-1}}_{J(y)} \ln\left(1-\frac{it}{it_c\pm y}\right) \nonumber
\end{align}
where we have substituted $y=e^{i\alpha}k^n$ and the Jacobian is given by $J(y)=e^{-i\alpha/n} y^{1/n-1}/n$. Since we are interested in potential cusps or divergences of $\tilde f(t)$, we consider the time derivative
\begin{equation}
    \label{fz3}
    \dot{\tilde{f}}(t) = \frac{i}{2\pi n}e^{-i\alpha/n}\int dy \frac{y^{1/n-1}}{it_c-it\pm y} \, .
\end{equation}
The integral is convergent for $n>1$ and we can evaluate it by choosing a proper branch cut and closing the integration contour. Using Eq.~\eqref{fz} we then obtain 
\begin{equation}
    \label{ldot}
    \dot{\tilde{l}}(t) = \pm \frac{1}{n}\sin\left(\frac{\pi-2\alpha}{2n}\right)|t-t_c|^{\frac{1}{n}-1} 
\end{equation}
where the sign refers to the cases $t<t_c$ and $t>t_c$, respectively. For the case of a constant density of Fisher zeroes, obtained by sending $n\to 1$, we find, in particular, that $\dot{\tilde{l}}(t)=\pm\cos(\alpha)$. I.e., in this case the return rate shows a cusp with finite derivatives whose shape is determined by the angle $\alpha$ at which the line of Fisher zeroes crosses the real time axis. For the more unusual case where the density of Fisher zeroes diverges along the curve of zeroes when approaching $t_c$
we have $J(y)\sim y^{1/n-1}$ with $n>1$ and a cusp with divergent derivatives is present. 

In some cases it is possible that the line of Fisher zeroes terminates exactly on the real time axis. Indeed this can be a generic feature for quenches which begin on a critical line~\cite{Ding2020,Wong2023b}. In this case the analysis above holds after mirroring the line of zeroes across the axis, resulting in only a difference of a factor of one half.

To summarize, there are discrete critical times $t_c$ in the one-dimensional case at which the return rate $l(t)$ shows a cusp whose properties are determined by the density of Fisher zeroes near $t_c$. 

\subsection{General considerations for dimensions \texorpdfstring{$d\geq 2$}{d>=2}}\label{sec:gen}
In more than one dimension, the Fisher zeroes will cross the real time axis in general not at isolated critical times $t_c$ but rather in entire critical regions. This can be understood, for example, for $d=2$ by the following considerations. In this case, the Fisher zeroes $z_p(k_x,k_y): \mathbb{R}^2\to\mathbb{C}$ can be parameterized as $z_p(k_x,k_y)=a_p(k_x,k_y)+i b_p(k_x,k_y)$ with real functions $a_p,b_p$. Potential singularities in $l(t)$ or its derivatives do occur when these Fisher zeroes cross the real time axis, $a_p(k_x,k_y)=0$. Here we can generally expect that there is a smooth curve $k_y=h(k_x)$ such that $a_p(k_x,h(k_x))=0$ leading to a critical region.

\begin{figure}[t!]
\includegraphics[width=0.99\columnwidth]{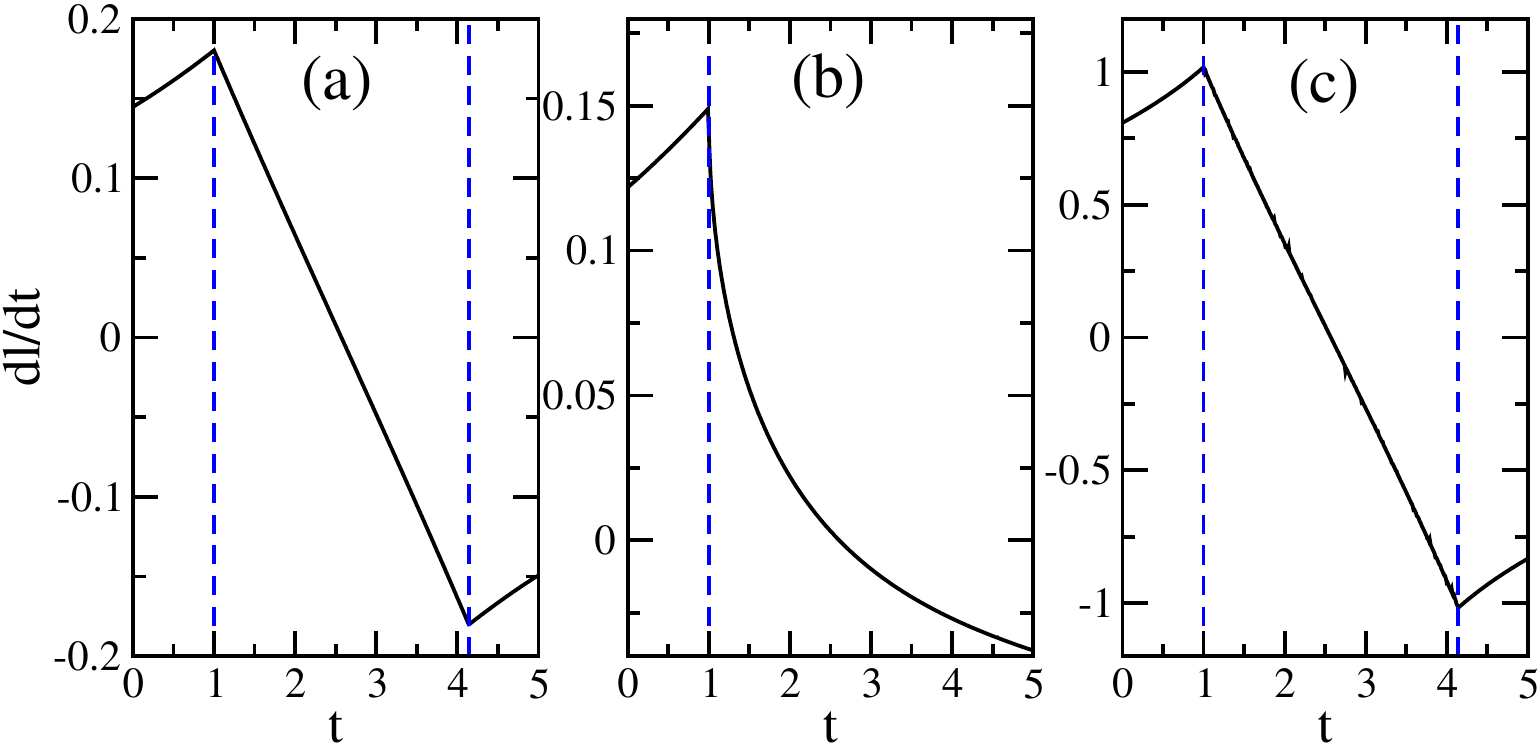}
\caption{$\dot{\tilde l}(t)$ for Fisher zeroes (a) $z(\vec{k})=it_c+ik_x+k_y$, (b) $z(\vec{k})=it_c+ik_x^2+k_y$, and (c) $z(\vec{k})=it_c+ik_x+k_y+k_z$ with $t_c=1$. All result in cusps at critical times. For a divergent density of Fisher zeroes as in example (b), the slope at the cusp from inside the critical region is infinite.} 
\label{Fig_analytic_example}
\end{figure}

The question is then what happens to the return rate $l(t)$ and its derivatives at the edges of such a critical region. We expect that this behaviour will be determined by the density of Fisher zeroes near the edge of the region and will be largely independent of the dimension $d\geq 2$. For a specific critical region with index $p$, we can write the singular part of $f(t)$ as
\begin{equation}
    \label{fzd1}
    \dot{\tilde f}(t)=\frac{i}{(2\pi)^d}\int\frac{d^dk}{z(\vec{k})-it} \, .
\end{equation}
The weakest singularities will occur when the density at the edge is constant. If the density of Fisher zeroes vanishes at the boundary, then $\dot{\tilde f}(t)$ will be smooth and non-analyticities will only show up in higher derivatives. We come back to this point below. Since the microscopic details should not affect the qualitative features of the singularities, we consider here a specific parametrization  of a constant density in two dimensions where $z(\vec{k})=it_c+ik_x+k_y$ with $k_x\in [0,\pi]$ resulting in a critical region $t_c<t<t_c+\pi$. For the singular part of the derivative of the return rate we obtain
\begin{align}
    \label{fzd2}
    \dot{\tilde f}(t)=&\frac{i}{(2\pi)^2}\int_0^\pi dk_x\int_{-\pi}^\pi dk_y  \frac{1}{it_c-it+ik_x+k_y} \nonumber \\
    =&\frac{i}{(2\pi)^2}\int_0^\pi \ln\left(\frac{it_c-it+ik_x+\pi}{it_c-it+ik_x-\pi}\right)  \, .
\end{align}
Here we can use the complex logarithm with the branch cut along the negative real time axis. We can then also perform the second integral giving the final result
\begin{align}
    \label{fzd3}
    \dot{\tilde l}(t)=&\textrm{Re}[\dot{\tilde f}(t)]=\frac{1}{4\pi^2}\bigg[\pi\ln\left(\frac{\pi^2+(t_c-t+\pi)^2}{\pi^2+(t_c-t)^2}\right) \\
    +&(t_c-t+\pi)\left\{\arg(-\pi+i(t_c-t+\pi))\right. \nonumber \\
    &\hspace{3cm}\left. -\arg(\pi+i(t_c-t+\pi))\right\} \nonumber \\
    +&(t_c-t)\left\{\arg(\pi+i(t_c-t))-\arg(-\pi+i(t_c-t))\right\} \bigg]. \nonumber
\end{align}
With the chosen branch cut we have $\arg(-\pi+i\epsilon)-\arg(-\pi-i\epsilon)=2\pi$, i.e., the last term in Eq.~\eqref{fzd3} jumps by $2\pi$ when crossing the critical time $t_c$. Due to the prefactor $t_c-t$ this leads to a cusp in $\dot{\tilde l}(t)$. Similarly, there is a jump for the argument function in the second line of Eq.~\eqref{fzd3} at $t_c+\pi$ resulting in a second cusp. This is shown in panel (a) of Fig.~\ref{Fig_analytic_example}.
We can also consider what happens if the density of Fisher zeroes diverges at the boundary of the critical region. An example is shown in panel (b) of Fig.~\ref{Fig_analytic_example} where $z(\vec{k})=it_c+ik_x^2+k_y$. We see that this also results in a cusp in $\dot{\tilde l}(t)$ but now with a divergent slope for $t\to t_c^+$. We expect this to be generic as well. 

For the three-dimensional case, nothing changes fundamentally. The Fisher zeroes, in general, still form regions in the two-dimensional complex time plane. They are now just parameterized by three instead of two momenta. In panel (c) of Fig.~\ref{Fig_analytic_example} we show an example where $z(\vec{k})=it_c+ik_x+k_y+k_z$ which leads to a constant density of states in the critical region. As in the two-dimensional case, there are then cusps in $\dot{\tilde l}(t)$ at the edges of this region with a finite slope. 

Generally speaking, we can define a density of Fisher zeroes $z_p({\bm k})$ in the complex plane $z=x+iy$ by 
\begin{align}\label{fishrhodef}
    \rho(z)
    =&\frac{1}{N}\sum_{p,{\bm k}}\delta(x-\textrm{Re}\,(z_p({\bm k})))\delta(y-\textrm{Im}\,(z_p({\bm k})))\\
    =&\sum_p\int \frac{d^d{\bm k}}{(2\pi)^d}\delta(x-\textrm{Re}\,(z_p({\bm k})))\delta(y-\textrm{Im}\,(z_p({\bm k}))). \nonumber
\end{align}
From now on we use $y\equiv t$ synonymously. Taking as an example the two-dimensional case where the Fisher zeroes in one region take the form $z(\vec{k})=it_c+ik_x^2+k_y$---which we have considered in panel (b) of Fig.~\ref{Fig_analytic_example}---we find for the density
\begin{equation}
    \label{dens_example_b}
    \rho(z)=\frac{1}{(2\pi)^2}\frac{1}{\sqrt{y-t_c}}\theta(y-t_c)\theta(\pi^2-y+t_c)\, .
\end{equation}
I.e., in this case the density shows a square root singularity at the critical time $t_c$. For every microscopic model we can, at least in principle, determine the Fisher zeroes $z_p({\bm k})$ and then use Eq.~\eqref{fishrhodef} to calculate their density in the complex plane. This will then tell us at which times cusps in $\dot{\tilde l}(t)$ will occur and what the shape of these cusps will be.

\subsection{Fisher zeroes in two dimensions}

We will now concentrate on studying Fisher zeroes and the related singular behaviour in derivatives of the return rate $l^{(m)}(t)$ for the two-dimensional case. If we change Eq.~\eqref{fishrhodef} to coordinates $x_p'=\textrm{Re}\,(z_p({\bm k}))$ and $y_p'=\textrm{Im}\,(z_p({\bm k}))$ we have 
\begin{equation}
    \rho(x,y)=\sum_n\int \frac{dx_p'dy'}{4\pi^2}\left|J_p(x_p',y')\right|\delta(x-x_p')\delta(y-y_p') \label{def_rhok}
\end{equation}
with the Jacobian determinant
\begin{equation}
\label{Jacobian}
    J_p(x',y')=\left|\frac{\partial (k_x, k_y)}{\partial(x_p',y_p')}\right|=\frac{1}{\left|\frac{\partial(x_p',y_p')}{\partial (k_x, k_y)}\right|}\, .
\end{equation}
For a given model, the calculation of this Jacobian determinant will thus allow to obtain the density of Fisher zeroes in the complex plane. The return rate is then given by
\begin{equation}\label{2sing}
    l(z)=-\int \frac{dx'dy'}{4\pi^2}\left|J(x',y')\right|\ln\left(1-\frac{z}{x'+iy'}\right).
\end{equation}

It is also possible in two dimensions to have an area of Fisher zeroes which terminates exactly on the real time axis at a single point. Indeed this behaviour is generic for the model in section \ref{sec:results} when quenching from a critical line. In that case we have the limiting result where the cusps at the beginning and end of the Fisher zeroes crossing the real time axis coalesce, and they are removed.

\subsubsection{Two-band models}
As already mentioned, dynamical quantum phase transitions do occur in topological matter of which the simplest examples are two-band models. Such models can be described by
\begin{equation}
\label{2band1}
\hh=\sum_{{\bm k}}\Psi^\dagger_{{\bm k}}\hc_{{\bm k}}\Psi_{{\bm k}}=\sum_{{\bm k}}\Psi^\dagger_{{\bm k}}{\bm d}_{{\bm k}}\cdot\vec{{\tau}}\Psi_{{\bm k}}
\end{equation}
where $\Psi^\dagger_j=(c_j^\dagger,c_j)$ with $c_{ j}^{(\dagger)}$ annihilating (creating) a spinless particle at site $j$. The Hamilton matrix in momentum space $\hc_{{\bm k}}$ can be written as $\hc_{{\bm k}}={\bm d}_{{\bm k}}\cdot\vec{\tau}$ where $\vec{d}_{{\bm k}}$ is a three dimensional vector and $\vec{\tau}$ is the vector of Pauli matrices.

If we assume that $\hh_1$ is particle-hole symmetric with eigenenergies $\pm\epsilon^1_{\bm k}$ then the Loschmidt amplitude takes the form~\cite{Vajna2015}
\begin{equation}\label{pbcloschmidt}
L(t)=\prod_{\bm k}\left[\cos[\epsilon^1_{\bm k}t]+i\hat{\mathbf{d}}^0_{\bm k}\cdot\hat{\mathbf{d}}^1_{\bm k}\sin[\epsilon^1_{\bm k}t]\right]\,.
\end{equation}
Here $\hat{\mathbf{d}}^{0,1}_k=\mathbf{d}^{0,1}_k/\left|\mathbf{d}^{0,1}_k\right|$ and $\mathbf{d}^{0,1}_k$ determine the Hamiltonians before and after the quench respectively. This result holds for all dimensions, for two-band models. Generalisations of such formulae to multi-band systems can also be found for specific cases~\cite{Maslowski2020}.

The Loschmidt amplitude has a product structure and thus vanishes if any of the factors is zero, which will happen at times
\begin{equation}
\label{zns}
z_p({\bm k})=\im t_p(\vec{k})=\frac{\im\pi}{\epsilon^1_{\bm k}}\left(p+\frac{1}{2}\right)-\frac{1}{\epsilon^1_{\bm k}}\tanh^{-1}\left[\hat{\mathbf{d}}^0_{\bm k}\cdot\hat{\mathbf{d}}^1_{\bm k}\right]\,. 
\end{equation}
Here $p$ is an integer. The Loschmidt echo $\mathcal{L}(t)$ thus vanishes at times
\begin{equation}
\label{critt}
t_p({\bm k}^*)=\frac{\pi}{\epsilon^1_{{\bm k}^*}}\left(p+\frac{1}{2}\right)
\end{equation}
where the ${\bm k}^*$ solve
\begin{equation}
\label{condition}
\hat{\mathbf{d}}^0_{{\bm k}^*}\cdot\hat{\mathbf{d}}^1_{{\bm k}^*}=0\,.
\end{equation}
As discussed more generally already in the previous section, this results in isolated points $k^*$ for one dimension, but for two dimensions one finds, in general, loops of critical momenta ${\bm k}^*$ in the Brillouin zone.

Writing again $x_p'=\textrm{Re}\,(z_p({\bm k}))$ and $y_p'=\textrm{Im}\,(z_p({\bm k}))$ with $z_p({\bm k})$ as defined in Eq.~\eqref{zns} we can calculate the Jacobian in Eq.~\eqref{Jacobian} and obtain
\begin{equation}\label{2bjacob}
    J_p(x_p',y_p')=-\frac{\pi^2(p+\frac{1}{2})^2}{(y_p')^3}\frac{1-\tanh^2\left[\frac{\pi x_p'}{y_p'}\left(p+\frac{1}{2}\right)\right]}{\frac{\partial\epsilon^1_{\bm k}}{\partial k_x}\frac{\partial \hat{\mathbf{d}}^0_{\bm k}\cdot\hat{\mathbf{d}}^1_{\bm k}}{\partial k_y}-\frac{\partial\epsilon^1_{\bm k}}{\partial k_y}\frac{\partial \hat{\mathbf{d}}^0_{\bm k}\cdot\hat{\mathbf{d}}^1_{\bm k}}{\partial k_x}}\,.
\end{equation}
We want to understand the possible divergences in the Jacobian which will determine where the density of Fisher zeroes in the complex plane is divergent. We note that $1/y_p'$ is regular. Divergences thus only occur if the Jacobian determinant in the denominator of Eq.~\eqref{2bjacob} vanishes, $|\partial(\epsilon^1_{\bm k},\hat{\mathbf{d}}^0_{\bm k}\cdot\hat{\mathbf{d}}^1_{\bm k})/\partial(k_x,k_y)|=0$. Now consider an area of Fisher zeroes crossing the real time axis. We then have a critical region $t_p=y_p'\in[y'_{\textrm{min}},y'_{\textrm{max}}]$ along the imaginary axis. According to Eq.~\eqref{critt}, for a fixed $p$ the energy $\epsilon^1_{\bm k}$ is at a local maximum or minimum at the boundaries of the interval. We can determine these local extrema under the constraint \eqref{condition} by using a Lagrange multiplier $\lambda$ leading to
\begin{equation}
    \label{Lagrange}
    f(k_x,k_y,\lambda)= \epsilon^1_{\bm k} + \lambda\,\hat{\mathbf{d}}^0_{{\bm k}}\cdot\hat{\mathbf{d}}^1_{{\bm k}} \, .
\end{equation}
From the condition for a local extremum $\nabla_{(k_x,k_y)}f(k_x,k_y)=0$ it then immediately follows that  $|\partial(\epsilon^1_{\bm k},\hat{\mathbf{d}}^0_{\bm k}\cdot\hat{\mathbf{d}}^1_{\bm k})/\partial(k_x,k_y)|=0$. To summarize, Eq.~\eqref{critt} shows that the energy has local extrema at the boundaries of the critical region and that at these boundaries the Jacobian in Eq.~\eqref{2bjacob} and therefore the density of Fisher zeroes $\rho(x,y)$ will diverge. From the discussion in the previous section we thus know that $\dot{\tilde{l}}(t)$ will show cusps at the boundaries of the critical regions with a slope which diverges when approaching the boundaries from inside the region.

It seems though as if no generic statement about the density of Fisher zeroes in two dimensions can be made beyond two-band models. What we can do in general, however, is to relate the density of zeroes with the derivative of the return rate. If we assume that $\rho(x,y)\sim \theta(y-t_c)/(y-t_c)^r$ then we find for the singular part of the derivative of the return rate
\begin{equation}
    \label{guess}
    \dot{\tilde{l}}(t)\sim i\int  \frac{\rho(x,y)dy\, dx}{x+\im (y-t)}\sim \alpha+\beta t+\gamma\theta(t-t_c)(t-t_c)^{-r+1} 
\end{equation}
for $0\leq r<1$ with constants $\alpha,\beta,\gamma$. I.e., we have a cusp in $\dot{\tilde{l}}(t)$ at $t=t_c$. If $r>0$, and the density of Fisher zeroes thus diverges at $t_c$, then $\ddot{\tilde{l}}(t)$ diverges for $t\to t_c^+$ leading to a cusp in $\dot{\tilde l}(t)$ at $t\to t_c^+$ with an infinite slope. We also note that for $r<0$, i.e.,~for the case where the density of Fisher zeroes vanishes at the boundary, there is no cusp in $\dot{\tilde l}(t)$. Instead, cusps in higher derivatives will be present.

\section{Results for specific models}\label{sec:results}

To illustrate our findings, we now discuss exemplary two-dimensional and three-dimensional topological models. Although our results should generically apply for both integrable and non-integrable models, it is not possible to obtain sufficiently large system sizes for an analysis of non-integrable models in two or three dimensions. State of the art techniques give lattice sizes of the order of $4\times4$~\cite{Peotta2021,Brange2022} which is well under what would be required to see DQPTs in the return rate derivative or the behaviour of the density of Fisher zeroes in the complex plane.

\subsection{Results in two dimensions}

\begin{figure}[t!]
\includegraphics[height=0.85\columnwidth]{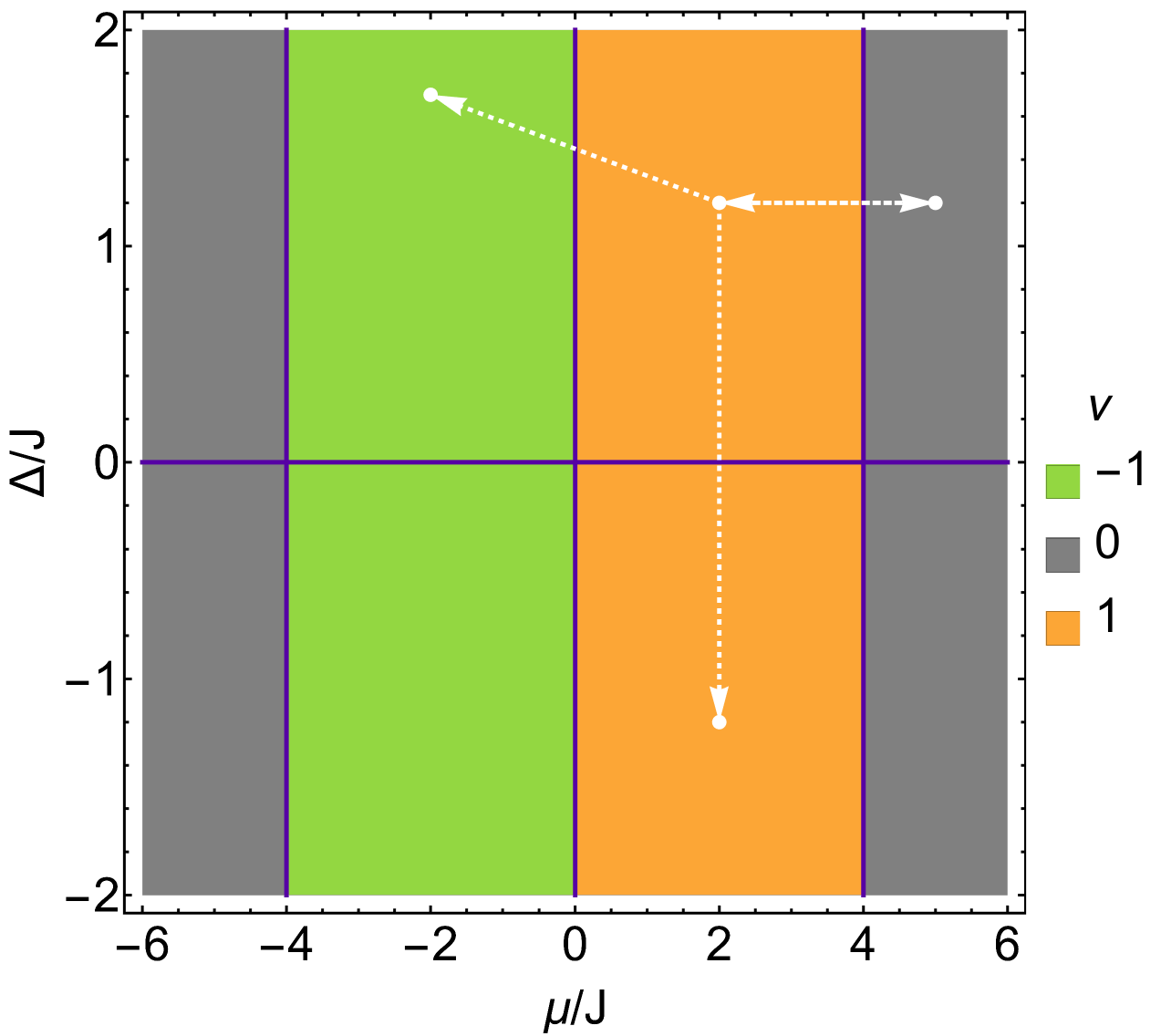}
\caption{The topological phase diagram of the two-dimensional Kitaev model, see Eq.~\eqref{kit_hamk}, showing the Chern number $\nu$. Bulk gap closings are solid purple lines. White dashed arrows show the quenches that are considered in this article, for parameters see Table~\ref{tab:quenches}.} 
\label{fig:phase}
\end{figure}

For the two-dimensional case, we take a generalisation of the Kitaev chain to a $p_x+ip_y$-wave topological superconductor~\cite{Kitaev2001,Sedlmayr2015b}. The two-dimensional Kitaev model is given by
\begin{equation}\label{kit_hamk}
    \hat{H}=
    \sum_{{\bm k}}\Psi^\dagger_{{\bm k}}\hc_{{\bm k}}\Psi_{{\bm k}}
    \equiv\sum_{{\bm k}}\Psi^\dagger_{{\bm k}}{\bm d}_{{\bm k}}\cdot\vec{{\bm\tau}}\Psi_{{\bm k}}\,,
\end{equation}
where
\begin{equation}\label{2d_d}
    \vec{d}_{\bm k}=
    \begin{pmatrix}
        -2\Delta\sin k_y \\
        -2\Delta\sin k_x \\
        -2J\cos k_x-2J\cos  k_y-\mu
    \end{pmatrix}.
\end{equation}
$\Psi^\dagger_{{\bm k}}=(c^\dagger_{{\bm k}},c_{{\bm k}})$ with $c_{{\bm k}}^{(\dagger)}$ annihilating (creating) a spinless particle with momentum ${{\bm k}}$. $J$ is the hopping strength and $\Delta$ the pairing strength with $\mu$ the chemical potential. $\vec{\tau}=({\tau}^x,{\tau}^y,{\tau}^z)^T$ are Pauli matrices representing particle-hole space. This model has eigenvalues $\pm\epsilon_{{\bm k}}$ given by
\begin{equation}
	\epsilon^2_{{\bm k}}=(2J\cos k_x+2J\cos k_y+\mu)^2+4\Delta^2(\sin^2k_x+\sin^2k_y).
\end{equation}
The phase diagram of the Kitaev model is straightforward to calculate using the Chern number, see App.~\ref{app:chern} and Fig.~\ref{fig:phase}, on which we also mark the quenches we consider. In Table \ref{tab:quenches}, the parameters used in the various quenches for the Kitaev model are given.

\begin{table}
    \begin{center}
    \begin{tabular}{|c|c|}
        \hline Parameters & Chern No.\\
        $(\mu/J,\Delta/J)$ & $\nu$   \\\hline
         $(5,1.2)$ & 0   \\\hline
         $(2,1.2)$ & 1   \\\hline
         $(-2,1.7)$ & -1   \\\hline
         $(2,-1.2)$ & 1   \\\hline
    \end{tabular}
    \caption{The points in the phase diagrams used for the different quenches of the Kitaev model, see Fig.~\ref{fig:phase} and Eq.~\eqref{kit_hamk}.}
    \label{tab:quenches}
    \end{center}
\end{table}

\begin{figure}
\includegraphics[width=0.45\columnwidth]{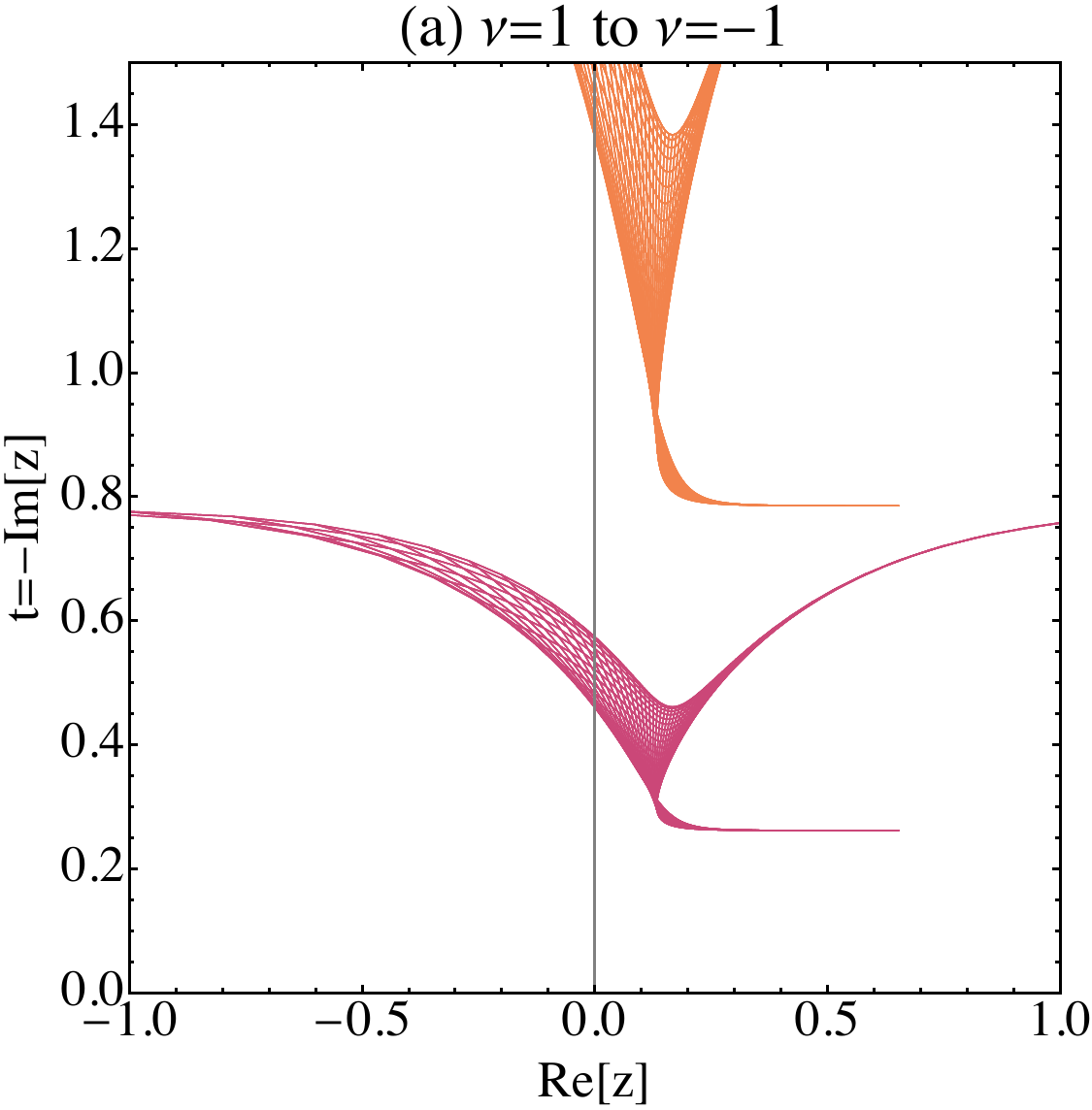}
\includegraphics[width=0.45\columnwidth]{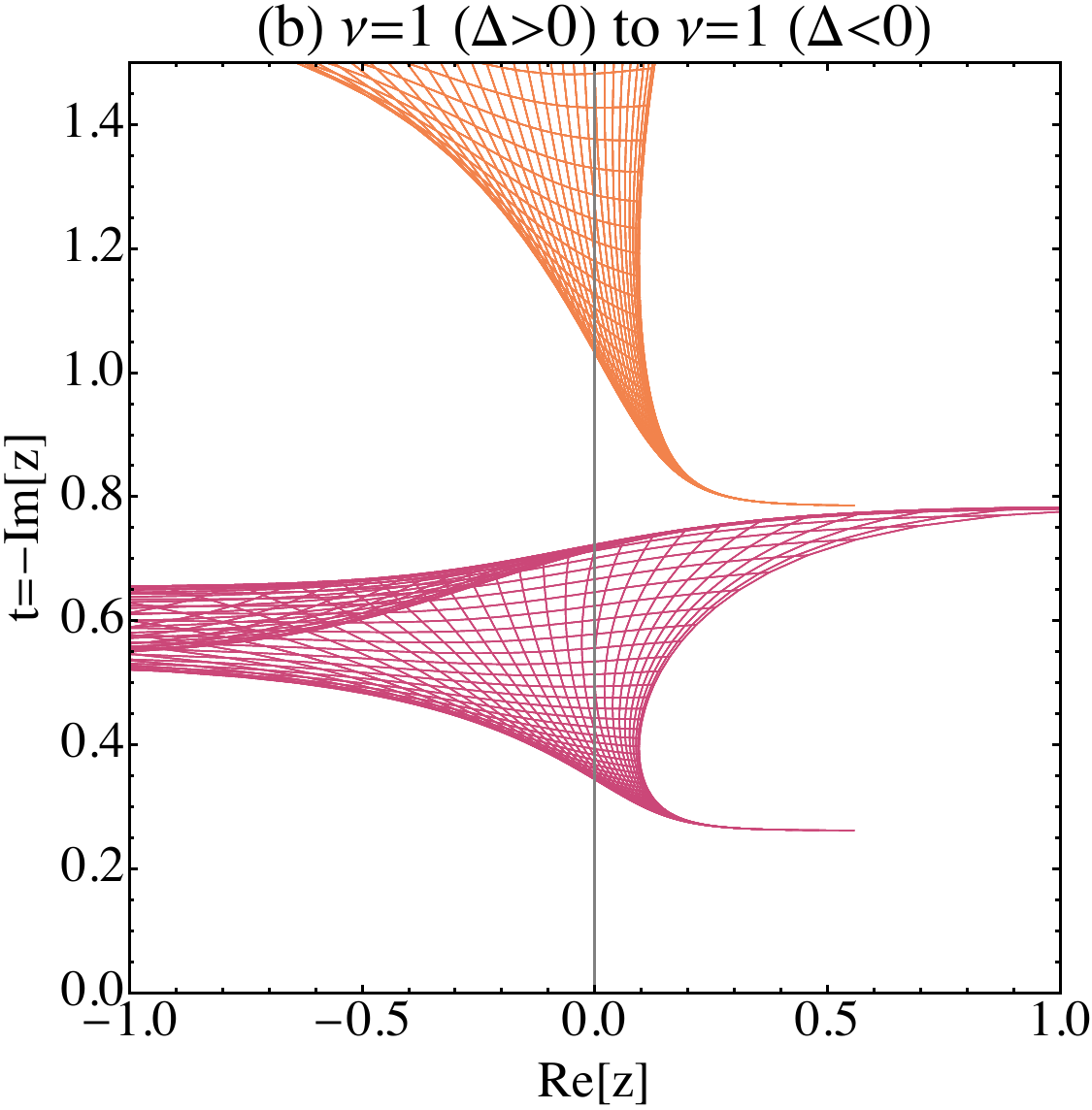}\\
\includegraphics[width=0.45\columnwidth]{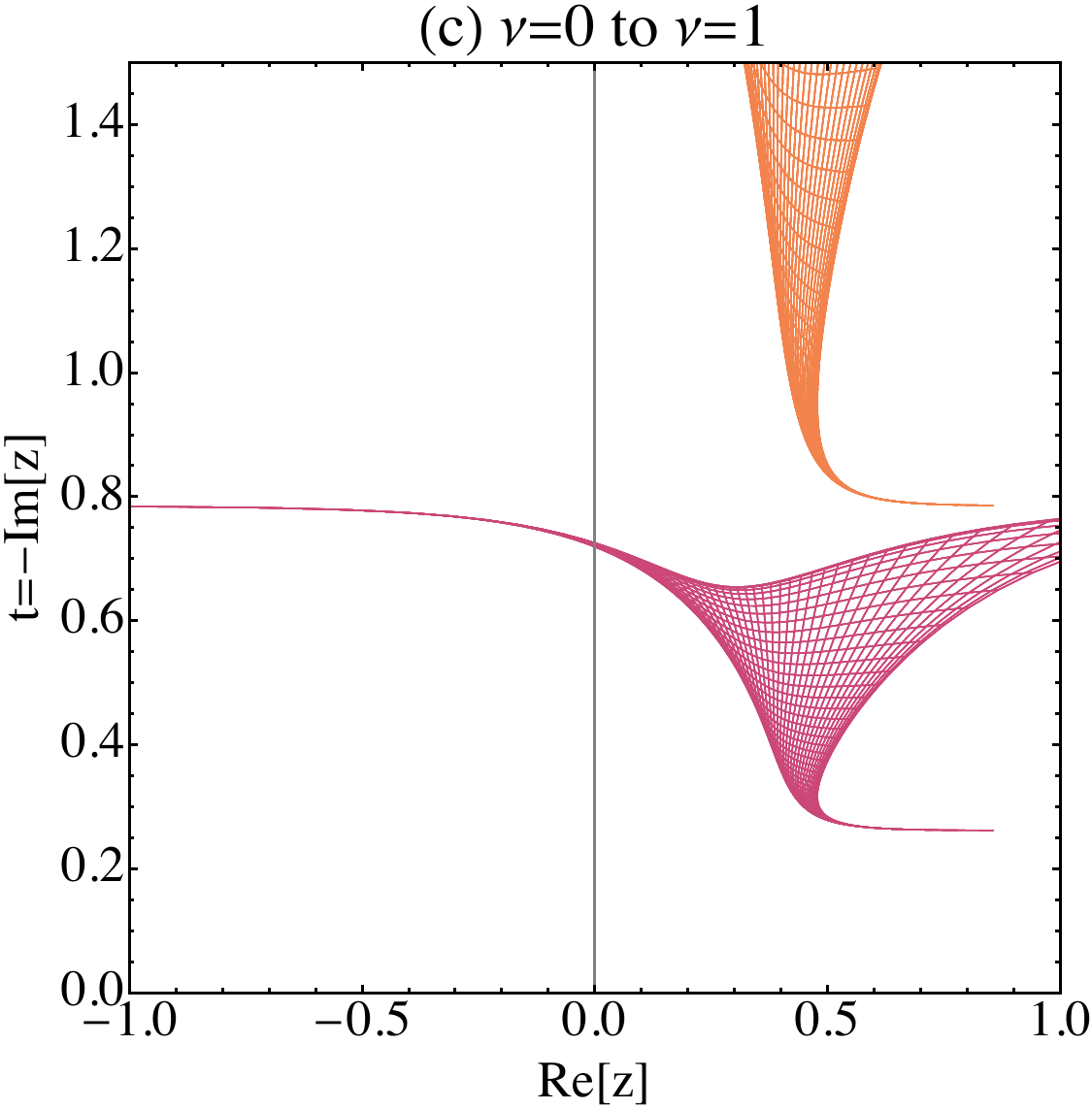}
\includegraphics[width=0.45\columnwidth]{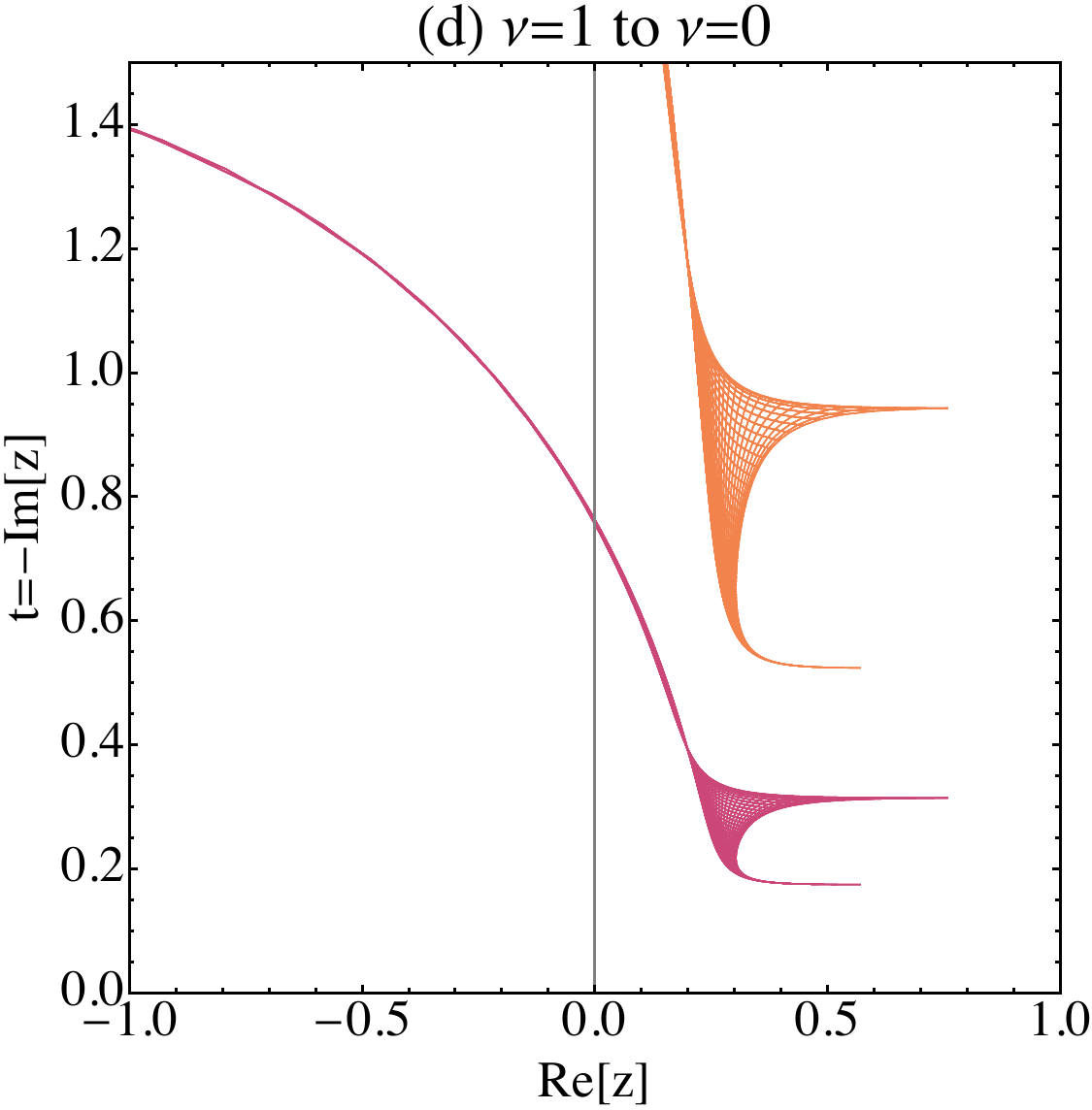}
\caption{Fisher zeroes for the Kitaev model quenches, see Table \ref{tab:quenches}. The critical regions corresponding to $p=0$ (red) and $p=1$ (orange) are shown. In all of these cases there are DQPTs caused by the Fisher zeroes crossing the real time axis, even for quenches within the same phase. The density of the Fisher zeroes diverges at the extremal times when crossing the real time axis. The system size is $N=100^2$.} 
\label{fig:fish}
\end{figure}

\begin{figure}
\includegraphics[width=0.45\columnwidth]{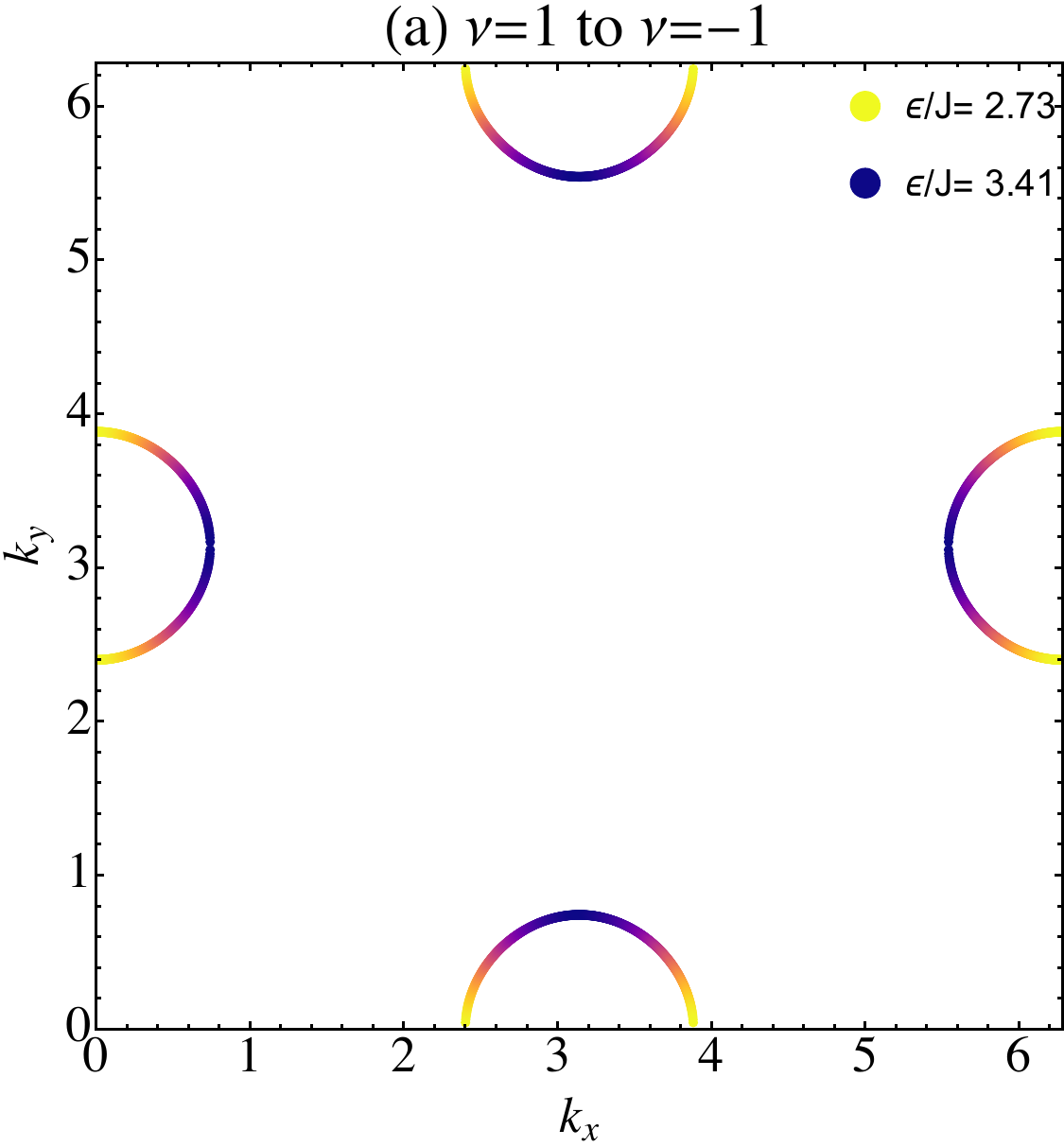}
\includegraphics[width=0.45\columnwidth]{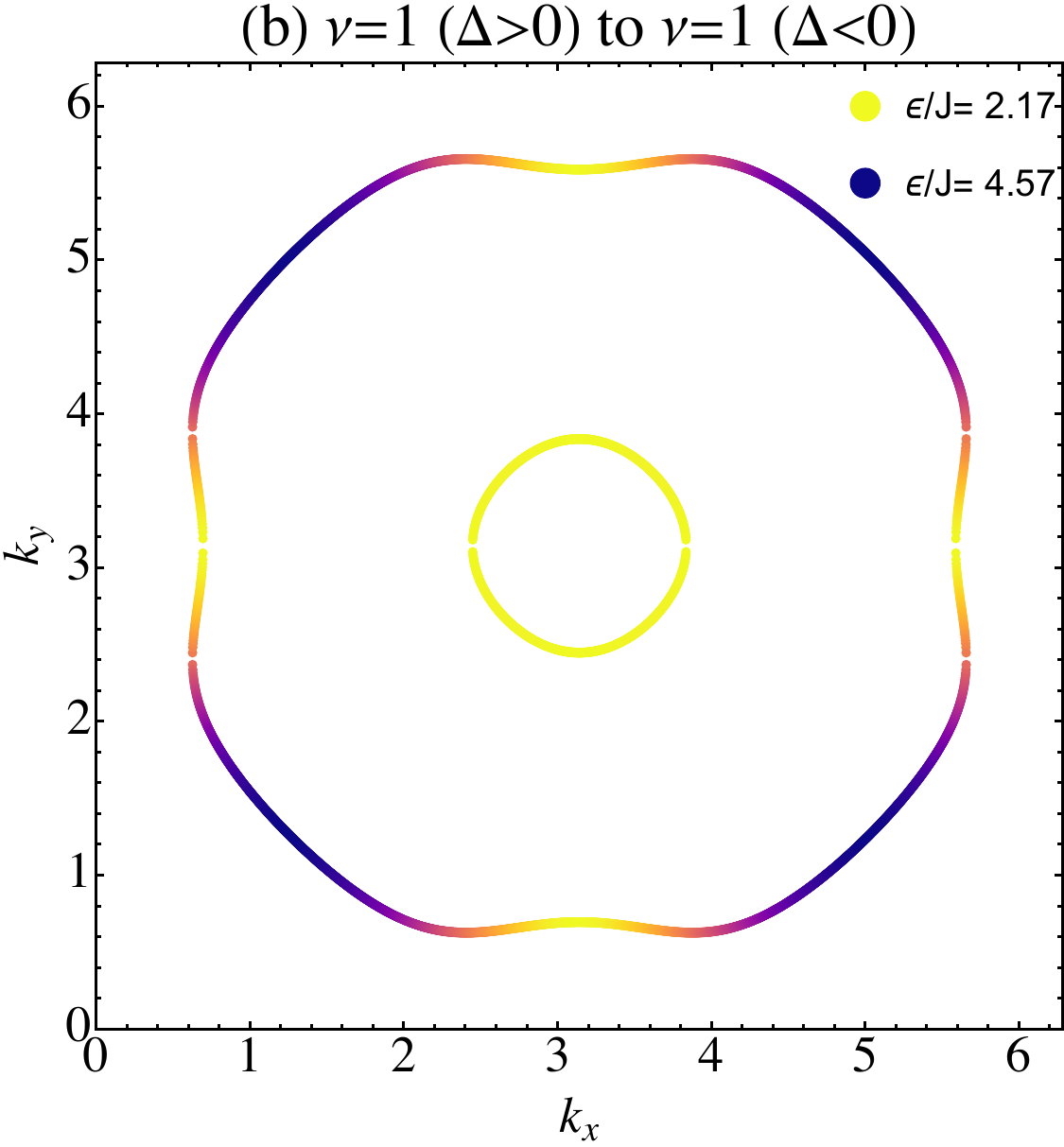}
\caption{Kitaev model: Critical momenta ${\bm k^*}$ for (a) a quench from $\nu=1$ to $\nu=-1$, and (b) from $\nu=1$ and $\Delta>0$ to $\nu=1$ and $\Delta<0$. The colour illustrates $\epsilon^1_{{\bm k}^*}$ with the extrema as indicated in the plot.  The system size is $N=4001^2$.} 
\label{fig:cm}
\end{figure}

We first consider the Fisher zeroes for the quenches in the Kitaev model. According to Eq.~\eqref{zns}, the Fisher zeroes come in different branches characterized by an integer $p$ and can be determined from the dispersion of the quench Hamiltonian and the $d$-vectors before and after the quench. The results are shown in Fig.~\ref{fig:fish}.
As expected for a two-band model in two dimensions, we find regions of Fisher zeroes with densities which diverge at the boundaries of these regions. The Loschmidt echo will become zero when the Fisher zeroes cross the real time axis  which happens, according to Eq.~\eqref{condition}, at points $\vec{k}^*$ in the Brillouin zone where the $d$-vectors are orthogonal to each other. In two dimensions, these critical momenta must form loops, and examples of such loops are shown in Fig.~\ref{fig:cm}.

\begin{figure*}
\includegraphics[height=0.45\columnwidth]{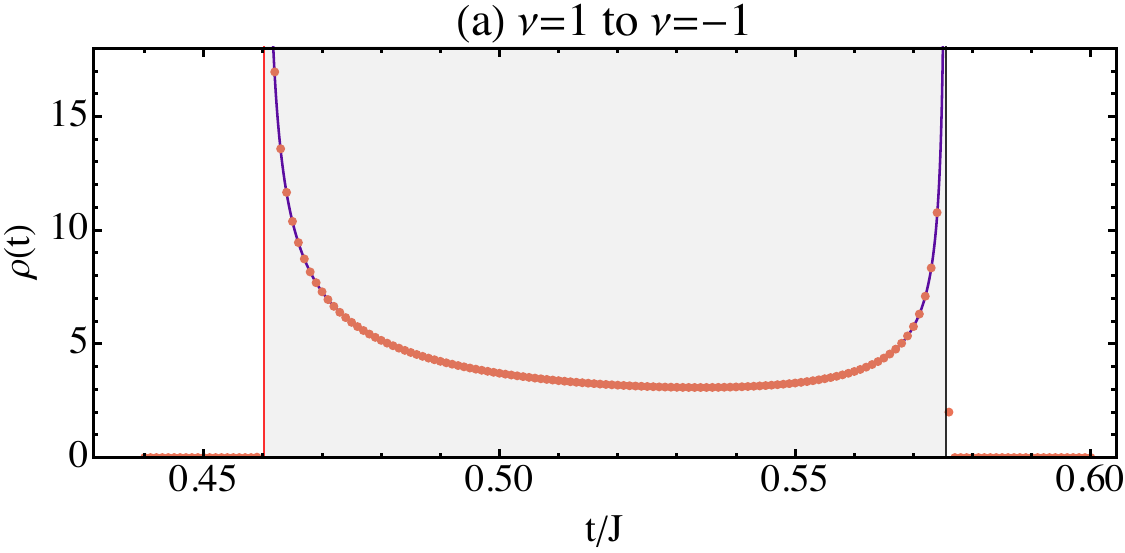}
\includegraphics[height=0.45\columnwidth]{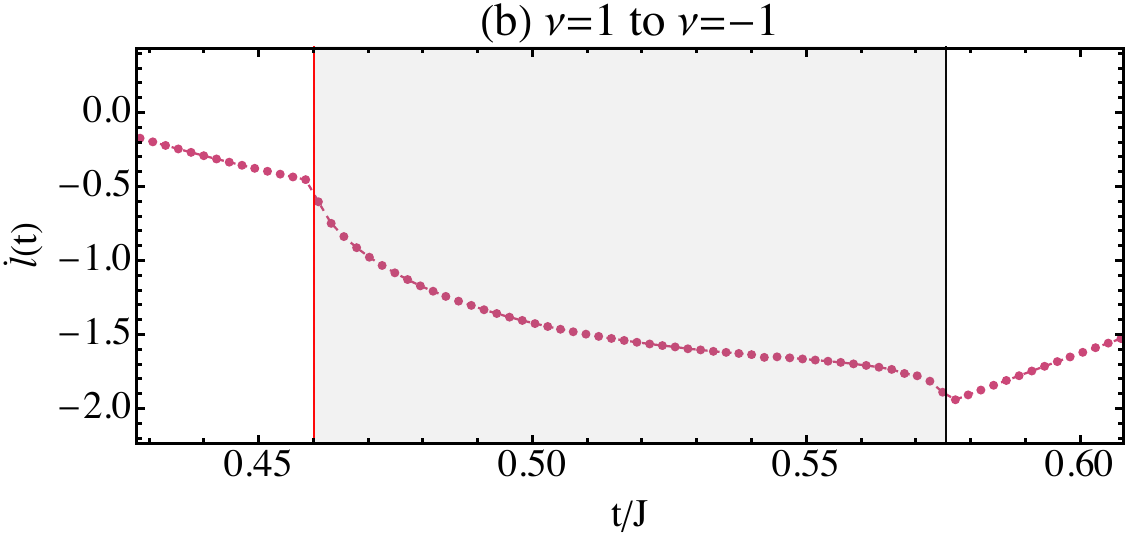}\\
\includegraphics[height=0.45\columnwidth]{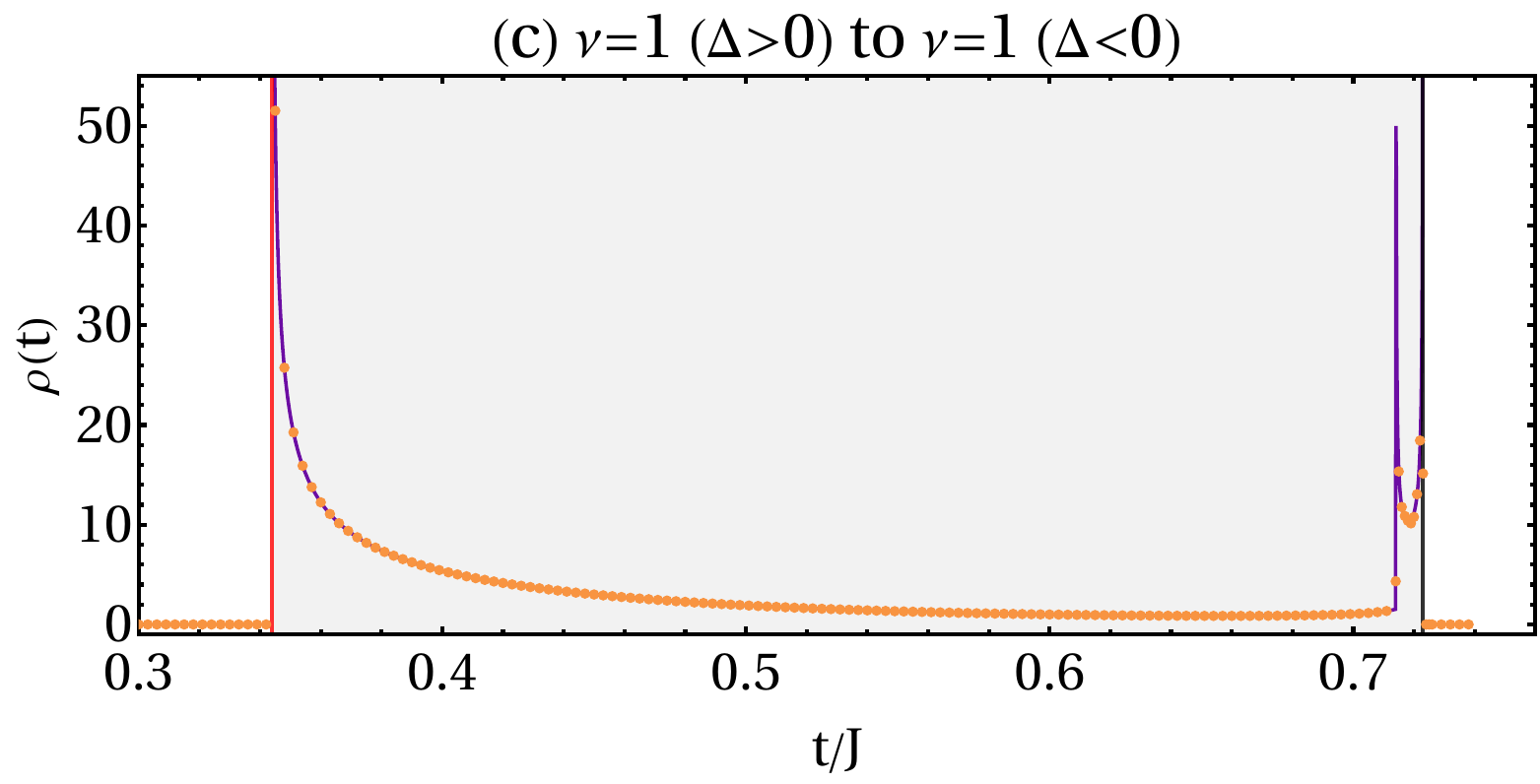}
\includegraphics[height=0.45\columnwidth]{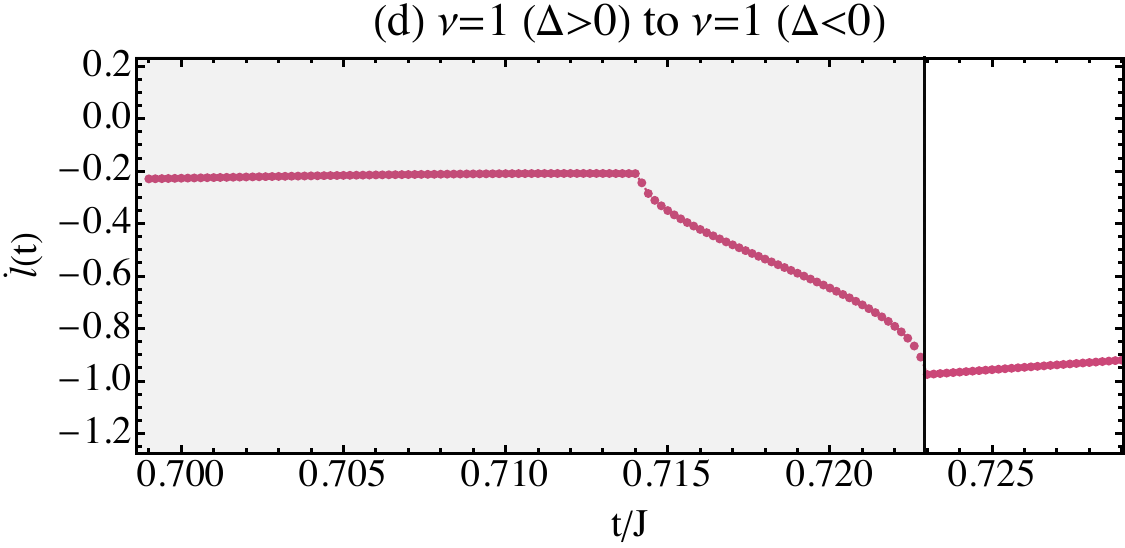}
\caption{Kitaev model: Density of Fisher zeroes along the real time axis, $\rho(t,0)$ (a,c), compared with the behaviour of the derivative of the bulk return rate $\dot{l}(t)$ (b,d). Quenches are as marked on the figures. For the density of Fisher zeroes the analytical result (solid lines), see App.~\ref{app:kit_det}, is compared to a numerical approximation (circles) calculated from the critical times, see Eq.~\eqref{critt}. The shaded regions are the critical regions. In the quench $\Delta\to-\Delta$, panels (c,d), there is additionally a divergence of the density {\it inside} the critical region, leading to an extra cusp in the return rate derivative, which can be seen in the zoom in panel (d). Return rates are calculated by numerical integration.} 
\label{fig:crit}
\end{figure*}

\begin{figure}
\includegraphics[height=0.45\columnwidth]{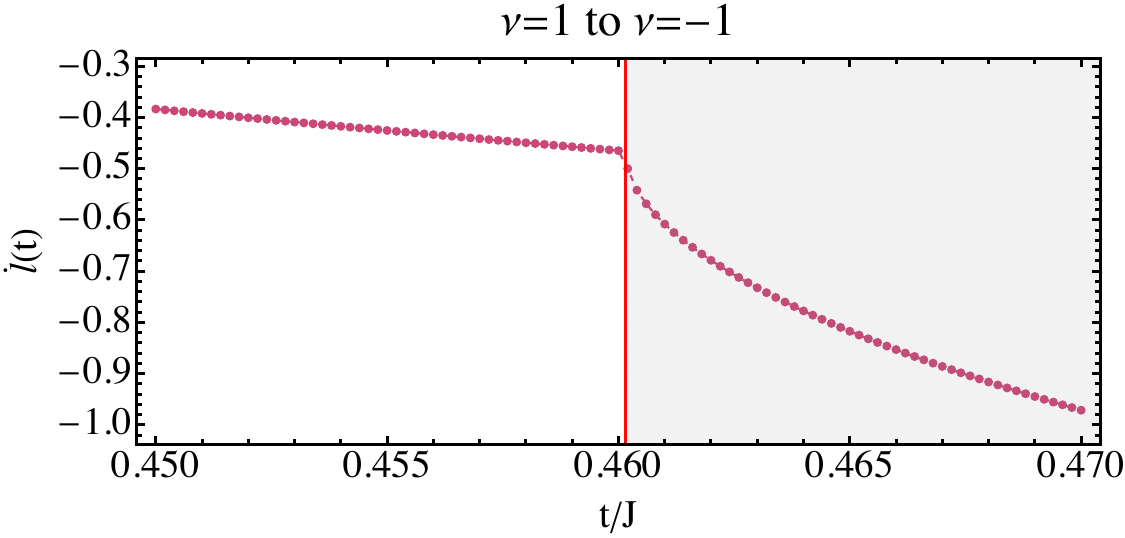}
\caption{Kitaev model: Zoom of Fig.~\ref{fig:crit}(b) showing that the cusp in $\dot l(t)$ has infinite slope for $t\to t_c^+$.} 
\label{fig:critz}
\end{figure}

For the Kitaev model, we can also analytically calculate the density of Fisher zeroes in the critical regions along the time axis $z=\im t$, see App.~\ref{app:kit_det} for more details. As examples, the results for two different quenches are shown in Fig.~\ref{fig:crit}.
At the boundaries of the critical regions we find that the density diverges as $\rho(t)\sim 1/\sqrt{|t-t_c|}$. We can also calculate the derivative of the return rate $\dot l(t)$ and find cusps at the boundaries of the critical regions, as predicted on general grounds in the previous section. In the second example, the quench does not change the winding number yet there are still cusps in $\dot l(t)$. Furthermore, this example also shows that additional cusps can occur inside the critical region if the density of Fisher zeroes has additional divergences, see appendix \ref{app:kit_det} for more details. The zoom in Fig.~\ref{fig:critz} shows that the square root divergence in the density leads to a cusp in $\dot l(t)$ with infinite slope for $t\to t_c^+$.

In App.~\ref{app:sq} we show results beyond a simple two-band model which demonstrate the same behaviour, i.e.,~cusps in $\dot{l}(t)$ at critical times. Finally it is interesting to note that in special cases the critical region can collapse down to a single critical time~\cite{Gulacsi2020,Maslowski2024}. In such a case the return rate itself once again shows a cusp as we have demonstrated earlier based on general grounds. 

\subsection{Results in three dimensions}

As an exemplary three-dimensional model we take a chiral topological insulator with three possible topological phases characterized by $\nu=0,1,2$ and with the usual Dirac surface states~\cite{Liu2023b}. The Hamiltonian density in momentum space is given by
\begin{equation}\label{3dham}
    \hc_{\bf k}=\vec{d}_{\bm k}\cdot\vec{{\bm\Gamma}}
\end{equation}
where
\begin{equation}
\label{3d_d}
    \vec{d}_{\bm k}=
    \begin{pmatrix}
        w+v\left[\cos k_x +\cos k_y+\cos k_z\right] \\
        v \sin[k_x] \\
        v \sin[k_y] \\
        v \sin[k_z]
    \end{pmatrix}
\end{equation}
with parameters $v,w$ and
\begin{equation}
    \vec{{\bm \Gamma}}=
    \begin{pmatrix}
        {\tau}^x\otimes{ \sigma}^x \\
        {\tau}^x\otimes{ \sigma}^y \\
        {\tau}^x\otimes{ \sigma}^z \\
        {\tau}^y\otimes{ \sigma}^0 
    \end{pmatrix}\,.
\end{equation}
The Hamiltonian can then be written as $\hat{H}=
\sum_{{\bm k}}\Psi^\dagger_{{\bm k}}\hc_{{\bm k}}\Psi_{{\bm k}}$ where $\Psi_{{\bm k}}$ is a four dimensional vector of annihilation operators. $\vec{\tau}=({\tau}^x,{\tau}^y,{\tau}^z)^T$ and $\vec{\sigma}=({\sigma}^x,{\sigma}^y,{\sigma}^z)^T$ are vectors of Pauli matrices typically taken to refer to the orbital and spin degrees of freedom respectively. Here their exact physical interpretation is not relevant.

This model has doubly degenerate eigenvalues given by
\begin{align}
    \varepsilon_{\bm k}^2=&3 v^2+w^2
    +2vw\left[\cos k_x +\cos k_y+\cos k_z\right]\\\nonumber&
    +2 v^2 \left[\cos k_x\cos k_z+\cos k_y\cos k_z+\cos k_x\cos k_y\right]
\end{align}
The phase diagram is quite simple \cite{Liu2023b}. If we take both $v$ and $w$ to be positive we find: $0<v<w: \nu=-2$, $w<v<3w: \nu=1$, and $3w<v: \nu=0$.

One can then find that for such a four-band model one in fact has
\begin{equation}\label{los3d}
L(t)=\prod_{\bm k}\left[\cos[\epsilon^1_{\bm k}t]+i\hat{\mathbf{d}}^0_{\bm k}\cdot\hat{\mathbf{d}}^1_{\bm k}\sin[\epsilon^1_{\bm k}t]\right]^2\, ,
\end{equation}
and
\begin{equation}\label{return3d}
l(t)=-\frac{1}{N}\sum_{\bm k}\left[\cos^2\epsilon^1_{\bm k}t+\left(\hat{\mathbf{d}}^0_{\bm k}\cdot\hat{\mathbf{d}}^1_{\bm k}\right)^2\sin^2\epsilon^1_{\bm k}t\right],
\end{equation}
a simple generalisation of the two-band case, though we stress that this is not generic behaviour for a four-band model.

\begin{figure}
    \centering
    \includegraphics[width=0.95\columnwidth]{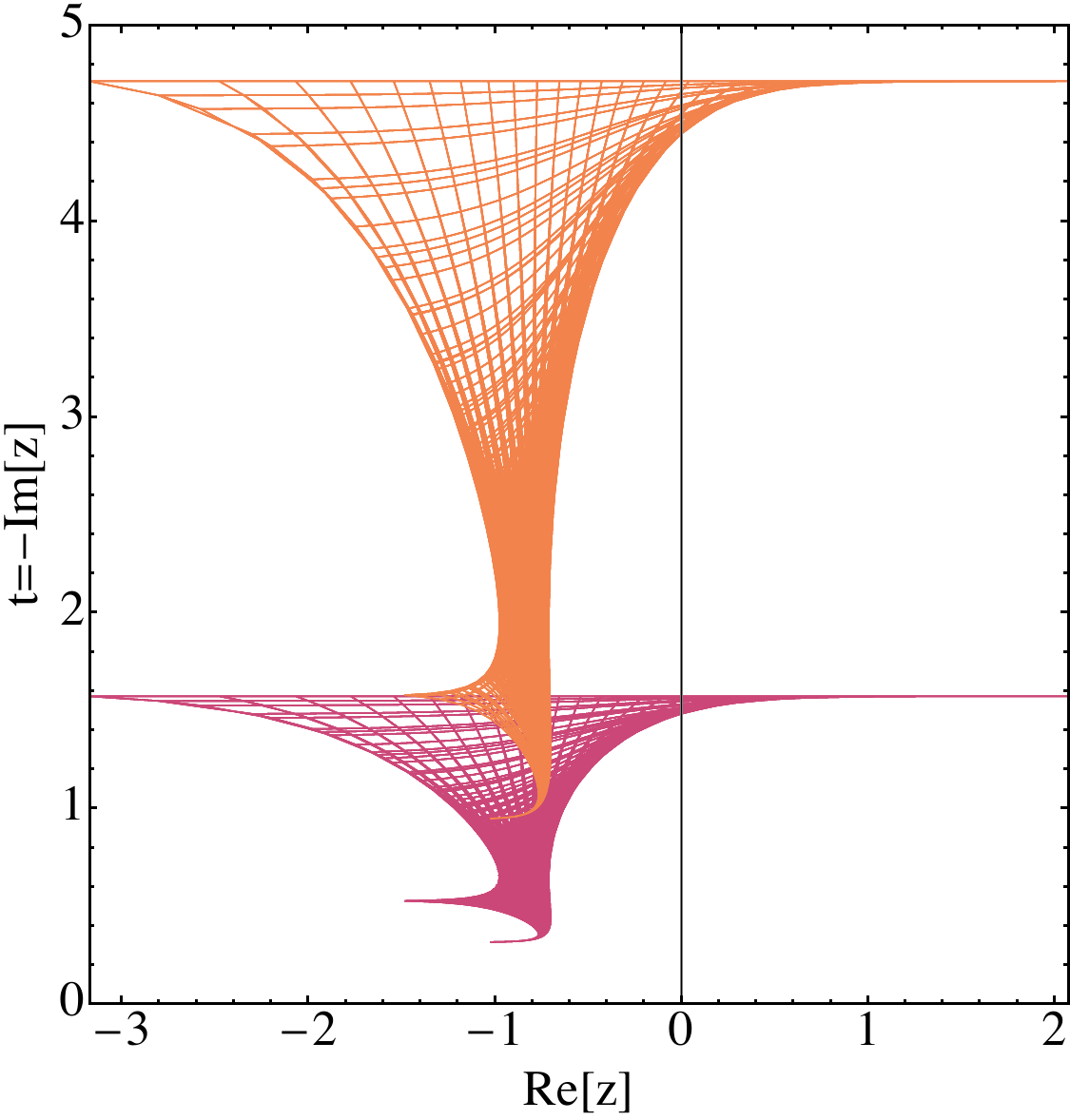}
    \caption{Quench in the three-dimensional model \eqref{3dham} with $\nu=0\to\nu=1$. Here we show the Fisher zeroes in the complex time plane at a system size $N=50^3$. The critical regions corresponding to $p=0$ (red) and $p=1$ (orange) are shown.}
    \label{fig:3D_Fisher}
\end{figure}

\begin{figure}
    \centering
    \includegraphics[width=0.95\columnwidth]{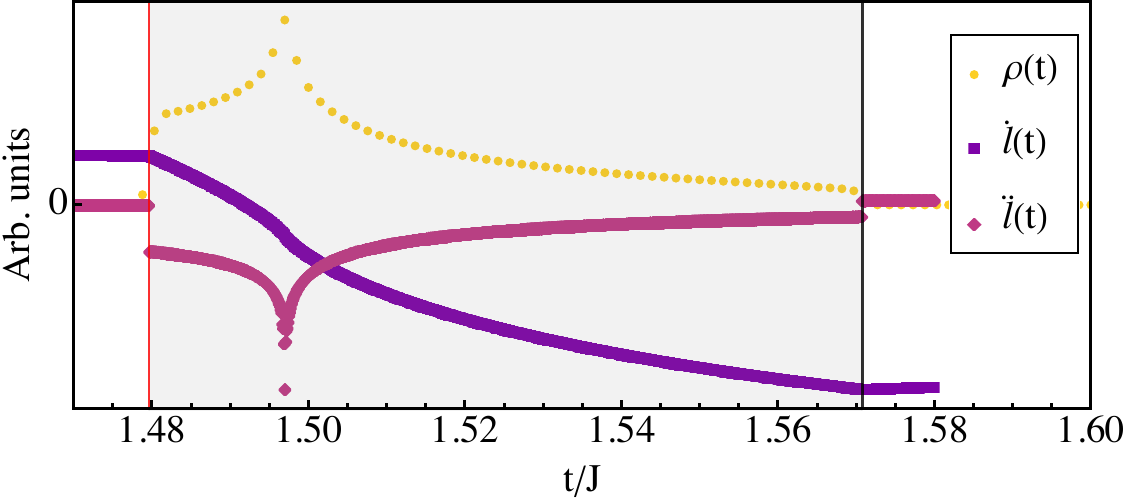}
    \caption{A comparison of $\dot{l}(t)$, $\ddot{l}(t)$, and the density of Fisher zeroes $\rho(t)$ for the quench in the three-dimensional model $\nu:0\to1$, see the main text for details. We focus here on the first critical region. The extent of the critical region is marked in gray and can be found by solving Eq.~\eqref{zns}. The density of Fisher zeroes, $\rho(t)$, is finite at the edges of the region, but shows an unexpected divergence inside the critical region. The jump in the density at the boundaries leads to a discontinuity in $\ddot l(t)$ and thus a cusp in $\dot l(t)$. For the divergence in $\rho$ we find a cusp in the second derivative. All quantities are scaled for ease of presentation. The second derivative was found using the VEGAS+ algorithm~\cite{Lepage2024}.}
    \label{fig:3D_DQPT}
\end{figure}

The Fisher zeroes are clearly still given by Eq.~\eqref{zns}. In Fig.~\ref{fig:3D_Fisher} we show the Fisher zeroes for a quench from the topologically trivial phase to the non-trivial phase with $\nu=1$, the quench parameters are $(w,v):(4,1)\to(2,1)$. We focus here on just this example. For $\textrm{Re}[z_p(\mathbf{k})]=0$ the critical momenta now form a surface in the three dimensional Brillouin zone. In Fig.~\ref{fig:3D_DQPT} we show the density of the Fisher zeroes along the real time line $t=-\textrm{Im}[z]$. The density is now constant at the edges of the critical regions, however divergences in the density are also visible.

In Fig.~\ref{fig:3D_DQPT} we compare the behaviour of the Fisher zero density to the return rate. The second derivative of the return rate shows a discontinuity at the edges of the critical region, corresponding to cusps in the first derivative $\dot{l}(t)$. Additionally, we see a divergence in $\rho(t)$ inside the critical region, which leads to a cusp in the second derivative of the return rate. 

\section{Conclusions}\label{sec:con}

In this article we have established a relation between the behaviour of Fisher zeroes in the complex time plane and DQPTs, i.e.,~singularities in the return rate or its derivatives, for one-, two-, and three-dimensional systems. 

For one-dimensional systems we have found that a line of Fisher zeroes with constant density crossing the real time axis leads to cusps in the return rate $l(t)$. The sharpness of these cusps is determined by the angle between the line of Fisher zeroes and the real time axis. We have also considered more exotic scenarios where the density of Fisher zeroes along the line diverges at the critical time $t_c$ where they cross the real time axis. In this case, the cusp can have a divergent slope at $t_c$.

In the two- and three-dimensional cases the Fisher zeroes form, in general, regions in the complex plane. We found that entering or exiting such a critical region leads to cusps in the derivative of the return rate $\dot l(t)$ as long as the density of Fisher zeroes is finite at $t_c$. If the density of Fisher zeroes diverges at the boundary, then the cusp in $\dot l(t)$ will have infinite slope when $t_c$ is approached from inside the critical region. For two-dimensional two-band models we have proven that the density does indeed always diverge at the boundaries of critical regions. Thus the scenario of a cusp with infinite slope on one side is realized in these cases.

To illustrate our results, we considered two examples. For a two-dimensional Kitaev model we were able to find explicit analytical formulae for the critical times and the scaling of the divergences of the density of Fisher zeroes. This is then related to the cusps in the derivative of the return rate which we obtained by numerical integration. We also showed that additional divergences can occur inside critical regions. The specific phenomenon that the density of Fisher zeroes shows a square root divergence at the edges of a critical region does not appear to be generic beyond two-band models and each general model must be considered individually. However, we have established a general, rigorous relation between the Fisher zero density and DQPTs. I.e., one can determine one from the other. We also note that we find DQPTs not only for quenches between phases with different Chern numbers but also for quenches within the same topological phase.

As a second example, we have considered a three-dimensional four-band topological insulator. For this model we derived analytical formulae for the Fisher zeroes and for the return rate. We find that for this model the density of Fisher zeroes is constant at the boundaries of the critical region in contrast to two-dimensional two-band models where this density diverges. This leads to cusps in $\dot l(t)$ at critical times $t_c$ with finite slopes. Inside the critical region we find a divergence of the density which leads to an additional critical time at which the second derivative of the return rate has a cusp.

\acknowledgments
This work was supported by the National Science Centre (NCN, Poland) under the grant 2019/35/B/ST3/03625. J.S.~acknowledges support by the German Research Council (DFG) via the Research Unit FOR 2316. J.S.~also acknowledges support by the National Science and Engineering Resource Council (NSERC) of Canada via the Discovery Grant program.

\appendix

\section{Chern number}\label{app:chern}

The Kitaev model considered in this work possesses only a particle-hole symmetry,
\begin{equation}
	\mathcal{C}\hc_{{\bm k}}\mathcal{C}=-\hc^*_{-{\bm k}},
\end{equation}
where $\mathcal{C}^2=1$. It is therefore in the D class of the topological periodic table with $\mathbb{Z}$ invariants in two dimensions~\cite{Ryu2010,Chiu2016}. The appropriate invariant is the Chern number, or equivalently the TKNN invariant~\cite{Thouless1982}, given by
\begin{equation}
\nu=\int \frac{d^2k}{2\pi} \sum_{j \ell}\mathcal{H}^{x}_{j \ell}\mathcal{H}^{y}_{\ell j}\frac{\textrm{sgn}\,[\epsilon_j]\Theta[-\epsilon_j\epsilon_\ell]}{(\epsilon_j-\epsilon_\ell)^2}\,.
\end{equation}
Here $S$ is the rotation to the energy eigenbasis of the Hamiltonian, i.e.,~$S^{-1}\hc_{{\bm k}}S$ is diagonal with eigenvalues $\epsilon_j$; then we define $S^{-1}\partial_{k_{x,y}}\hc_{{\bm k}}S\equiv\hc^{x,y}$. The momentum is denoted by ${\bf k}=(k_x,k_y)$. 

\section{Explicit formulae for the Fisher zero density for the Kitaev model}\label{app:kit_det}

Here we give an explicit formula for the density of Fisher zeroes for the Kitaev model \eqref{kit_hamk}, calculated from Eqs.~\eqref{2bjacob} for the branch $p=0$. Following some manipulation we find that for a quench from $(\mu_0,\Delta_0)$ to $(\mu_1,\Delta_1)$:
\begin{widetext}
\begin{align}
\rho(t)&=\sum_{s\in\{-1,1\}}\frac{(\mu_1+2 u^s_t)^2 \left|\Delta_1 (\mu_0+2 u^s_t)-\Delta_0(\mu_1+2 u^s_t)\right|^3 \sqrt{-4\gamma_t}}{\pi ^3 \left| \Delta_0^2 \Delta_1 \left(\Delta_1 (\mu_0+2 u^s_t)-(2\Delta_0-\Delta_1)(\mu_1+2 u^s_t)\right) \right|
\sqrt{4-\left(u^s_t\right)^2+2\gamma_t} \sqrt{\left(\left(u^s_t\right)^2-\gamma_t\right)^2-4\left(u^s_t\right)^2}},  \label{rho_exact}\\
\gamma_t&=\frac{(\mu_0+2 u^s_t) (\mu_1+2 u^s_t)}{4\Delta_0 \Delta_1},\\
u^s_t&=\frac{\left(\mu_1^2-\frac{\pi^2}{4t^2}\right) \Delta_0-\mu_0 \mu_1 \Delta_1}{(\mu_0+\mu_1) \Delta_1-2 \mu_1 \Delta_0+s\sqrt{\frac{\pi^2}{t^2} \Delta_0 (\Delta_0-\Delta_1)+(\mu_0-\mu_1)^2 \Delta_1^2}}. \label{ut}
\end{align}
\end{widetext}
We note that the above expression must be combined with a condition that the solutions for the momenta have physical values. From this one can quickly demonstrate by expansion that generically $\rho(t,0)\sim(t-t_c)^{-\frac{1}{2}}$ near a critical time for $t>t_c$ for this model.

From this expression we can also extract the critical times, see Eq.~\eqref{critt}. The critical interval can be determined by an analysis of extrema of $\epsilon^1_{\bm k}$. The condition for a local extremum, see Eq.~\eqref{Lagrange}, can be written as
\begin{equation}
    \sin k_x \cs (k_x,k_y,\lambda) =0\textrm{ and }  \sin k_y \cs (k_y,k_x,\lambda) =0\,, \label{extr_cnd}
\end{equation}
where
\begin{align}
    \cs(k_x,k_y,\lambda) =& \lambda  (\mu_0 + \mu_1)+ 2  \mu_1
    +4(1+\lambda)\cos k_y
    \\\nonumber&+ 
     4\left(1-\Delta_1^2+\lambda-\Delta_0
\Delta_1  \lambda \right)\cos k_x  \label{extr_cnd_in}
\end{align}
This leads to the following possible cases:
\begin{enumerate}
    \item First we can have
    \begin{equation}\label{extr_cnd_in}
        \cs(k_x,k_y,\lambda) = 0\,,\quad \cs(k_y,k_x,\lambda) = 0\,.
    \end{equation}
    Eq.~\eqref{extr_cnd_in} requires that $\cos k_x = \cos k_y$ which provides an expression for $\cos k_x$
    in terms of $\lambda$. Then the constraint $\hat{\mathbf{d}}^0_{{\bm k}}\cdot\hat{\mathbf{d}}^1_{{\bm k}}=0$ leads to a quadratic equation for $\lambda$. Finally, the critical times are
    \begin{widetext}
    \begin{align}
    t_{1,2}=&\frac{\pi }{2}\left|\Delta_0 \Delta_1-2\right|
    \left(16 \Delta _1 \left(\Delta_1 \left(\Delta_0^2+2\right)-\left(\Delta_1^2+2\right)\Delta_0\right)
    \phantom{\sqrt{\Delta_0^2}}\right. \nonumber\\&
    -\left(\Delta _1^2-2\right) \mu _0^2-\left(\Delta_0 \Delta_1 \left(\Delta _1^2-4\right)+4\right) \mu_1 \mu_0+\left(\Delta_0^2 \Delta_1^2-\Delta _1^2-2
   \Delta_0 \Delta_1+2\right) \mu_1^2 \nonumber \\
   &\left.  \phantom{\{} \pm\left|\left(\Delta_1^2-2\right) \mu _0+\left(\Delta _1^2-2 \Delta _0 \Delta_1+2\right) \mu _1\right| \sqrt{16\Delta_0^2 \Delta_1^2 +2 \Delta_0 \Delta_1\left(\mu_1 \mu _0-16\right) +\left(\mu_0-\mu_1\right)^2} \right)^{-\frac{1}{2}} .
   \end{align}
    \item The second possibility is 
    \begin{equation}
        \sin k_x = 0\,,\quad \cs(k_y,k_x,\lambda) = 0\,.
    \end{equation}
    The first equation implies two possibilities $\cos k_x=\pm1\equiv\xi$, which yields two solutions for $\cos k_y$ in terms of $\lambda$ coming from the second equation. Each of these solutions when applied to $\hat{\mathbf{d}}^0_{{\bm k}}\cdot\hat{\mathbf{d}}^1_{{\bm k}}=0$ gives two solutions for $\lambda$ and then two solutions for the critical time. Together we have four solutions, two for each sign of $\cos k_x$. These are
    \begin{align}
    t^\xi_{1,2}=&\frac{\pi }{\sqrt{2}}\left|\Delta _0 \Delta _1-1\right|
    \left( 16  \Delta_0  (\Delta_0 - \Delta_1) \Delta_1^2 
    - \left(4 \xi \Delta_1 \left( \Delta_1 + \Delta_0 (-2 + \Delta_1^2) \right) 
    - (1- \Delta_1^2) \mu_0  \right) \mu_0 \phantom{\sqrt{\Delta_0^2}} \right. \nonumber \\
    &- 2 (1+ \Delta_0  \Delta_1  (-2 + \Delta_1^2) ) \mu_0  \mu_1 
    -\left(4\xi\Delta_1^2 \left(1-2\Delta_0^2+\Delta_0\Delta_1\right)-
    \left(1-2\Delta_0\Delta_1+2\Delta_0^2\Delta_1^2-\Delta_1^2\right)\mu_1\right)\mu_1 \nonumber \\
    &\pm \left| 4\xi (\Delta_0-\Delta_1)\Delta_1 + \left(1 - \Delta_1^2 \right)\mu_0 
    - (1 - 2\Delta_0\Delta_1 + \Delta_1^2) \mu_1 \right| \nonumber \\
    &\left. \times \sqrt{16\Delta_0^2 \Delta_1^2 
    +4 \Delta _0 \Delta _1\left(\mu_0 \left(\mu_1+ 2\xi \right) + 2\xi \mu_1 \right) 
    +\left(\mu _0-\mu _1\right)^2} \right)^{-\frac{1}{2}} .
    \end{align}
    \end{widetext}
    The same can be obtained for $k_x\leftrightarrow k_y$.
    \item The final solution is for
    \begin{equation}
        \sin k_x = \sin k_y = 0\,.
    \end{equation}
    In this case the condition $\hat{\mathbf{d}}^0_{{\bm k}}\cdot\hat{\mathbf{d}}^1_{{\bm k}}=0$ can be satisfied only for $\cos k_x = \cos k_y = \pm 1$ which is only possible if $\mu_0=\mp 4$ or for $\cos k_x = - \cos k_y = \pm 1$ with $\mu_0=0$. These lead to critical times
    \begin{equation}
        t_c = \frac{\pi}{2|\mu_1\pm 4|}\quad \textrm{or}\quad t_c = \frac{\pi}{2|\mu_1|}
    \end{equation}
    accordingly. These are quenches which start from the critical point, $\mu_0=\mp 4$, or cross the real time axis at one point, $\mu_0=0$.
\end{enumerate}

\section{Further results for the Fisher zeroes and return rate for the Kitaev model}\label{app:kit_rr}

\begin{figure}[!t]
\hspace{0.35cm}\includegraphics[width=0.725\columnwidth]{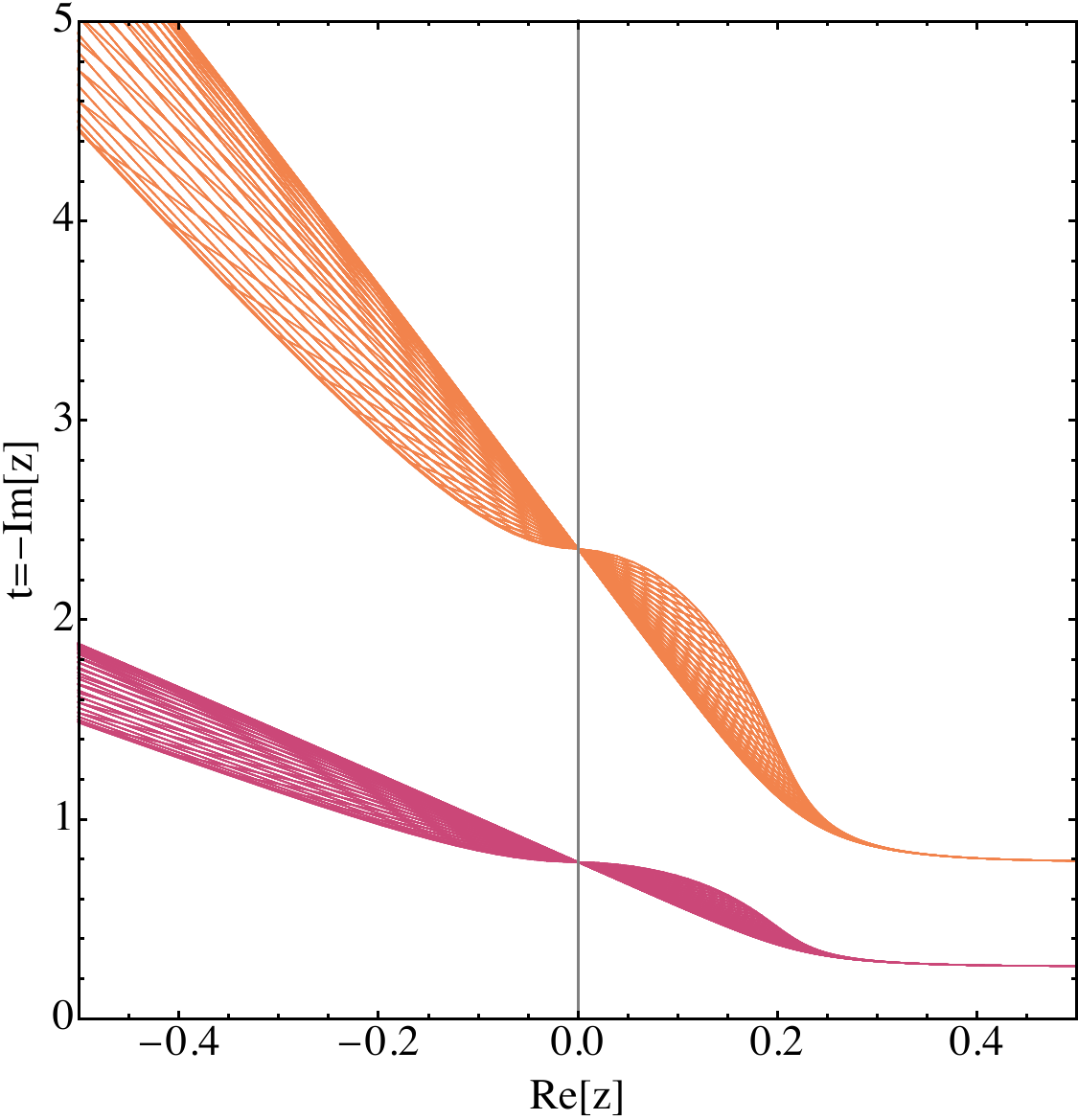}\\
\includegraphics[width=0.8\columnwidth]{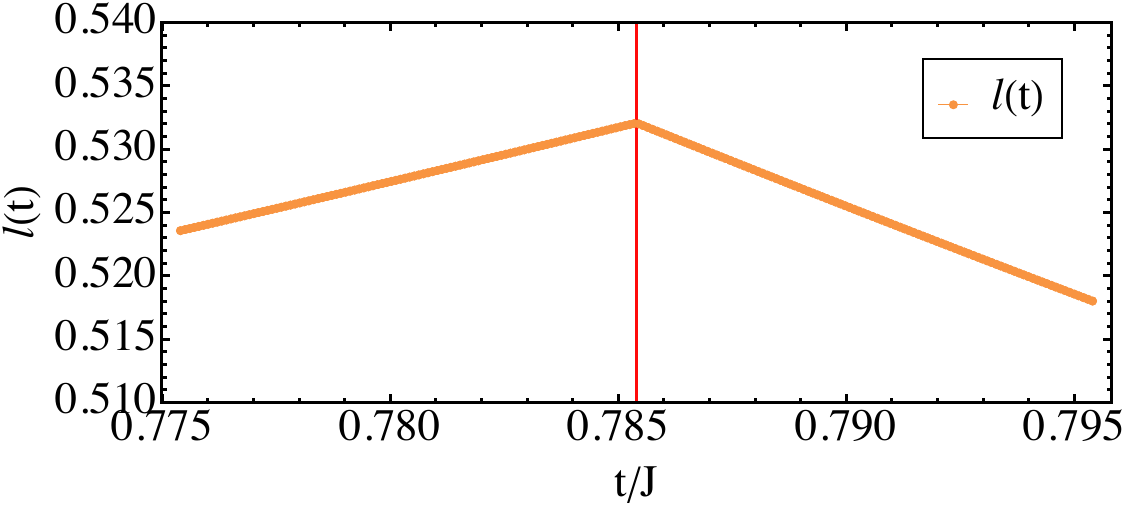}
\caption{Fisher zeroes and the return rate and  for a finely tuned quench of the Kitaev model, where $(\mu,\Delta):(0,-1)\to(2,0)$. In this case all Fisher zeroes which cross the real time axis do so at a single point, resulting in a cusp in the return rate itself. The momentum sums are performed by numerical integration and the Fisher zeroes are calculated for a lattice of size $N=100^2$.} 
\label{fig:rr_crit_1}
\end{figure}

\begin{figure}[!t]
\hspace{0.75cm}\includegraphics[width=0.725\columnwidth]{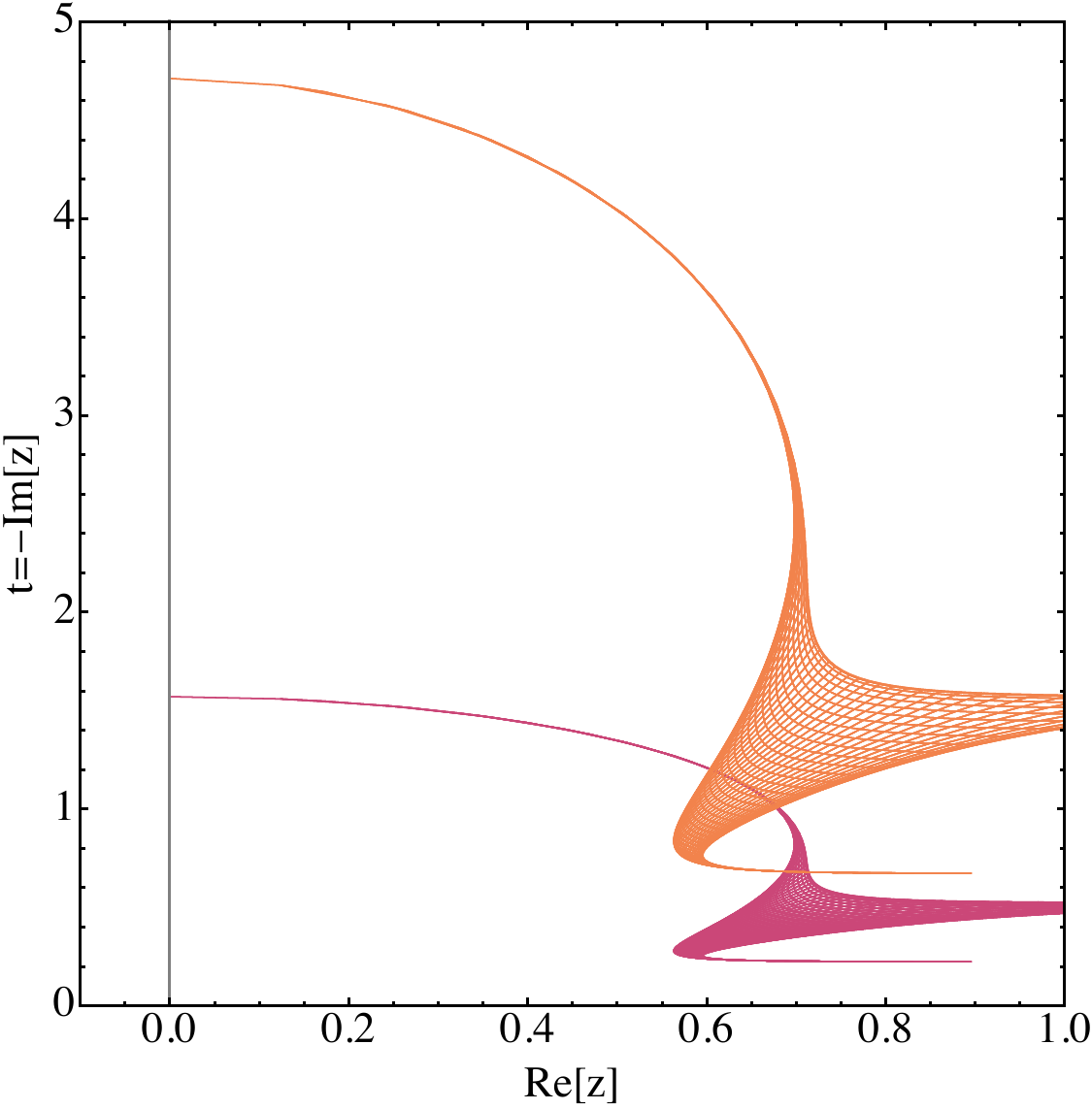}\\
\includegraphics[width=0.8\columnwidth]{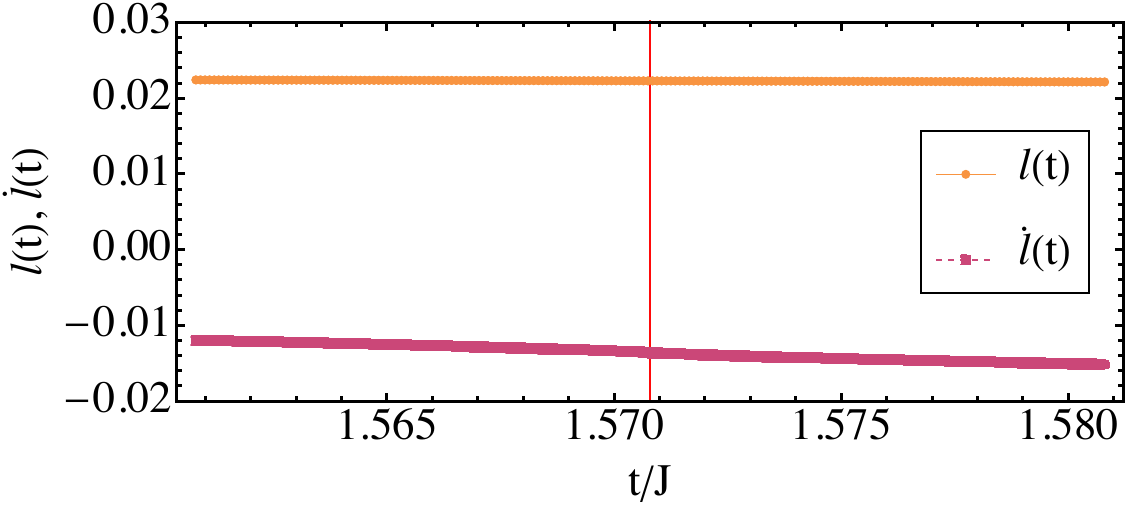}
\caption{Fisher zeroes, and the return rate and its derivative, for a quench of the Kitaev model, where $(\mu,\Delta):(4,1.2)\to(3,1.2)$. I.e.,~this is a quench from the critical line for which Fisher zeroes ending on the real time axis are a generic feature for this model. One can see that this is the limiting case at which the cusps for the DQPTs vanish. The momentum sums are performed by numerical integration and the Fisher zeroes are calculated for a lattice of size $N=100^2$.} 
\label{fig:rr_crit_2}
\end{figure}

Here we wish to show some of the more unusual cases that can occur. In Fig.~\ref{fig:rr_crit_1}, the Fisher zeroes all cross the real time axis at a single time resulting in a cusp in the return rate as in the one-dimensional case. Another interesting case are quenches starting from the critical line for which the Fisher zeroes in this specific model always end exactly on the real time axis, see Fig.~\ref{fig:rr_crit_2}. In this limiting case the cusps in the return rate vanish.

\begin{figure*}
\includegraphics[height=0.45\columnwidth]{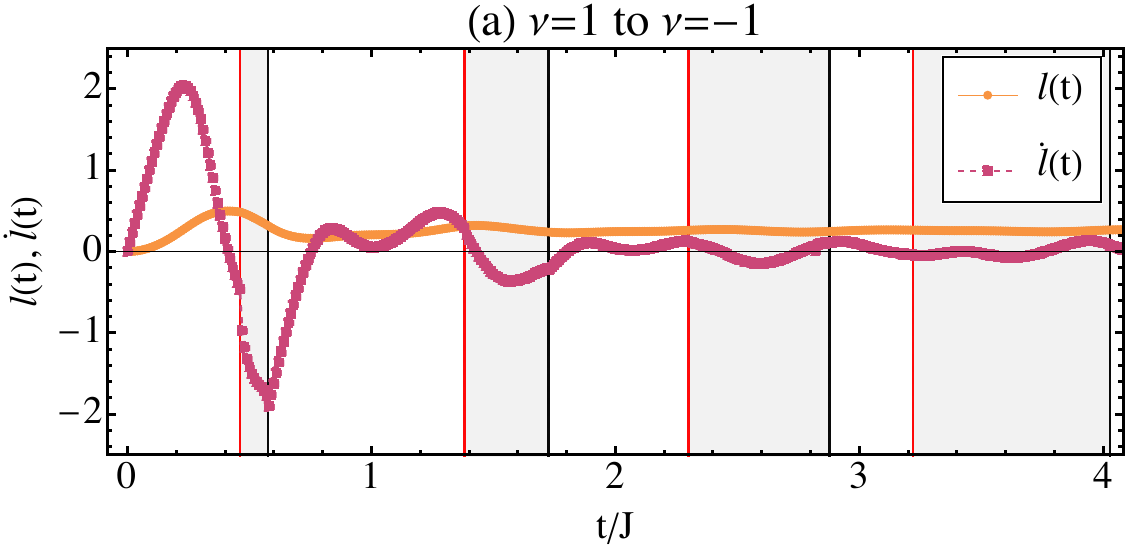}
\includegraphics[height=0.45\columnwidth]{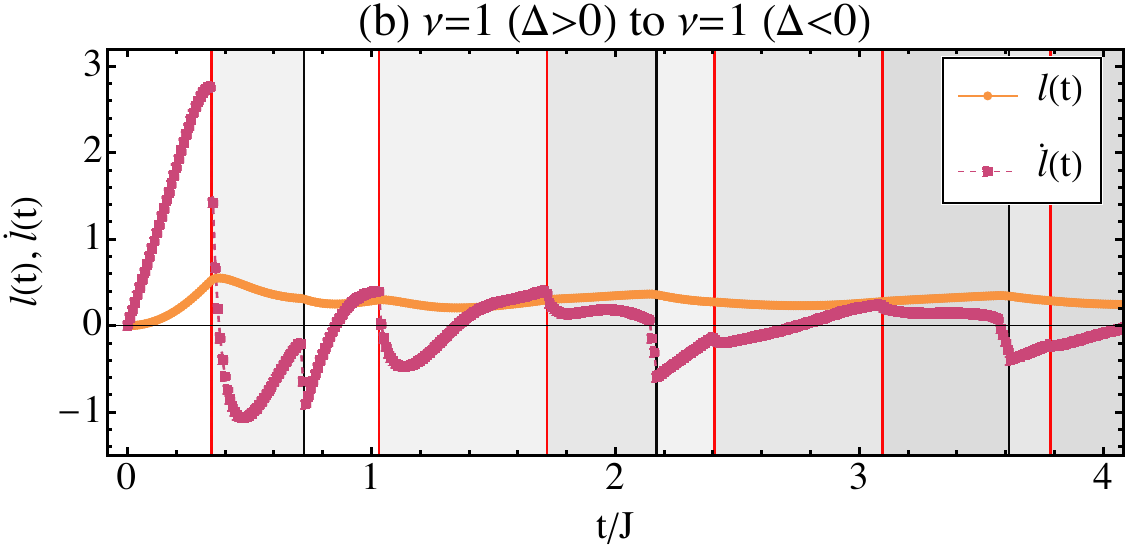}\\
\includegraphics[height=0.45\columnwidth]{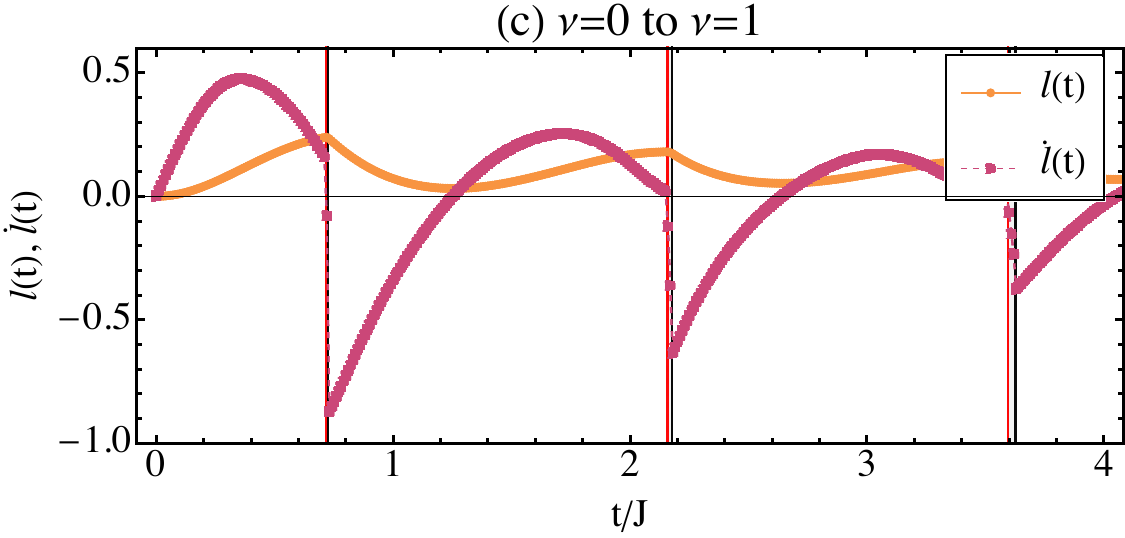}
\includegraphics[height=0.45\columnwidth]{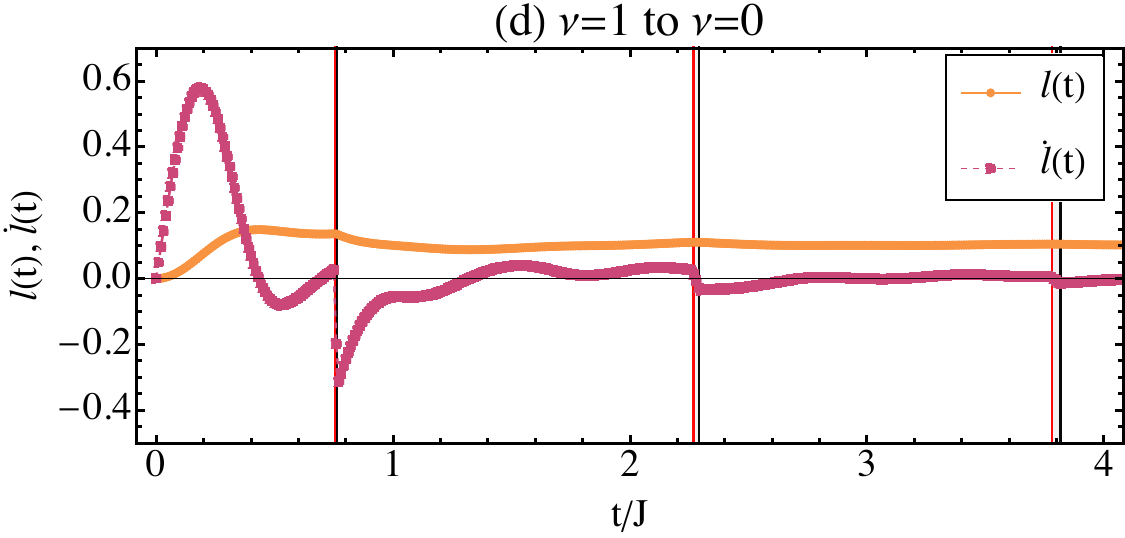}
\caption{The return rate and its derivative for the quenches of the Kitaev model, see table \ref{tab:quenches}. Critical regions are analytical results and shown shaded in grey. Discontinuities for $\dot{l}(t)$ are clearly visible, and $l(t)$ remains smooth. For some quenches it can be seen that the critical regions soon start to overlap with each other. The momentum sums are performed essentially in the thermodynamic limit as numerical integrals.} 
\label{fig:rr}
\end{figure*}

In Fig.~\ref{fig:rr}, results for all the four different quenches in table \ref{tab:quenches} are shown. While the return rate itself is always smooth, the first derivative does show cusps. We also note that the structure of $\dot l(t)$ can quickly become quite complex to disentangle because critical regions from different ``branches'' $p$ can start to overlap, which will become worse with increasing time as critical regions become progressively longer.

\section{Beyond two-band models}\label{app:sq}

In this appendix, we show an example for a two dimensional system which is no longer a simple two-band model. We take the generalisation of the two-dimensional Kitaev model on a square lattice to the spinful case, which is given by the Hamiltonian density
\begin{align}
\hc_{\bm k}=&
-2J(\cos k_x +\cos k_y){\bm\tau}^z -\mu{\bm\tau}^z-\Delta{\bm\sigma}^x\\\nonumber&
+2\alpha\sin k_x{\bm\tau}^z{\bm\sigma}^y-2\alpha\sin k_y {\bm\tau}^z{\bm\sigma}^x
-B{\bm\sigma}^z,
\end{align}
where ${\bm\tau}^j$ are Pauli matrices for the particle-hole subspace and ${\bm\sigma}^j$ are Pauli matrices for the spin subspace. In the interests of brevity a tensor product is implied but not written, and identity matrices are also not written explicitly. For the parameters we have the chemical potential $\mu$, hopping strength $J$, spin-orbit coupling $\alpha$, a Zeeman field $B$, and finally $\Delta$ is the s-wave pairing strength. The full Hamiltonian is $\hat{H}=
\sum_{{\bm k}}\Psi^\dagger_{{\bm k}}\hc_{{\bm k}}\Psi_{{\bm k}}$ where $\Psi_{{\bm k}}=(c_{\bm{k},\uparrow},c_{\bm{k},\downarrow},c^\dagger_{\bm{k},\downarrow},-c^\dagger_{\bm{k},\uparrow})^T$. We focus here on a single example of a quench from $\nu=0$ to $\nu=1$ with parameters $(\mu,\Delta,\alpha,B):(2, 0.4, 0.4, 1)\to(2, 0.4, 0.4, 3)$.

As we no longer have a two-band system we can no longer derive a simple analytical formula for the Loschmidt echo, the Fisher zeroes, or the return rate. Instead we rely on the alternative expression for the Loschmidt amplitude~\cite{Levitov1996,Klich2003,Rossini2007,Sedlmayr2018,Maslowski2020}
\begin{equation}\label{rle}
    L(t)=\prod_{\bm k}\det{\bm M}_{\bm k}(t)\equiv\prod_{\bm k}\det\left[1-\mathbf{\C}_{\bm k}+\mathbf{\C}_{\bm k}e^{it \mathcal{H}_{\bm k}}\right],
\end{equation}
where we have defined the correlation matrix as $\left[\C_{\bm k}\right]_{ij}=\langle\Psi_0|\Psi^\dagger_i\Psi_j|\Psi_0\rangle$ and $\mathcal{H}_{\bm k}$ is taken to refer to the time evolving Hamiltonian density. The Loschmidt amplitude is therefore given by the eigenvalues of ${\bm M}_{\bm k}(t)$ and we will denote by convention the smallest of these eigenvalues at a time $t$ as $|\lambda_0(t)|$. One can easily calculate the expressions
\begin{equation}\label{bulkrrd}
\dot{l}(t)=-\frac{1}{N}\sum_{\bm k}\tr\dot{\bm M}_{\bm k}(t){\bm M}_{\bm k}^{-1}(t)
\end{equation}
and
\begin{equation}\label{bulkrrdd}
\ddot{l}(t)=-\frac{1}{N}\sum_{\bm k}\tr\left[\ddot{\bm M}_{\bm k}(t){\bm M}_{\bm k}^{-1}(t)-\dot{\bm M}^2_{\bm k}(t){\bm M}_{\bm k}^{-2}(t)\right],
\end{equation}
which can then be evaluated numerically using also expressions for the derivatives of the Loschmidt matrix ${\bm M}_{\bm k}(t)$.

\begin{figure}
    \centering
    \includegraphics[width=0.95\columnwidth]{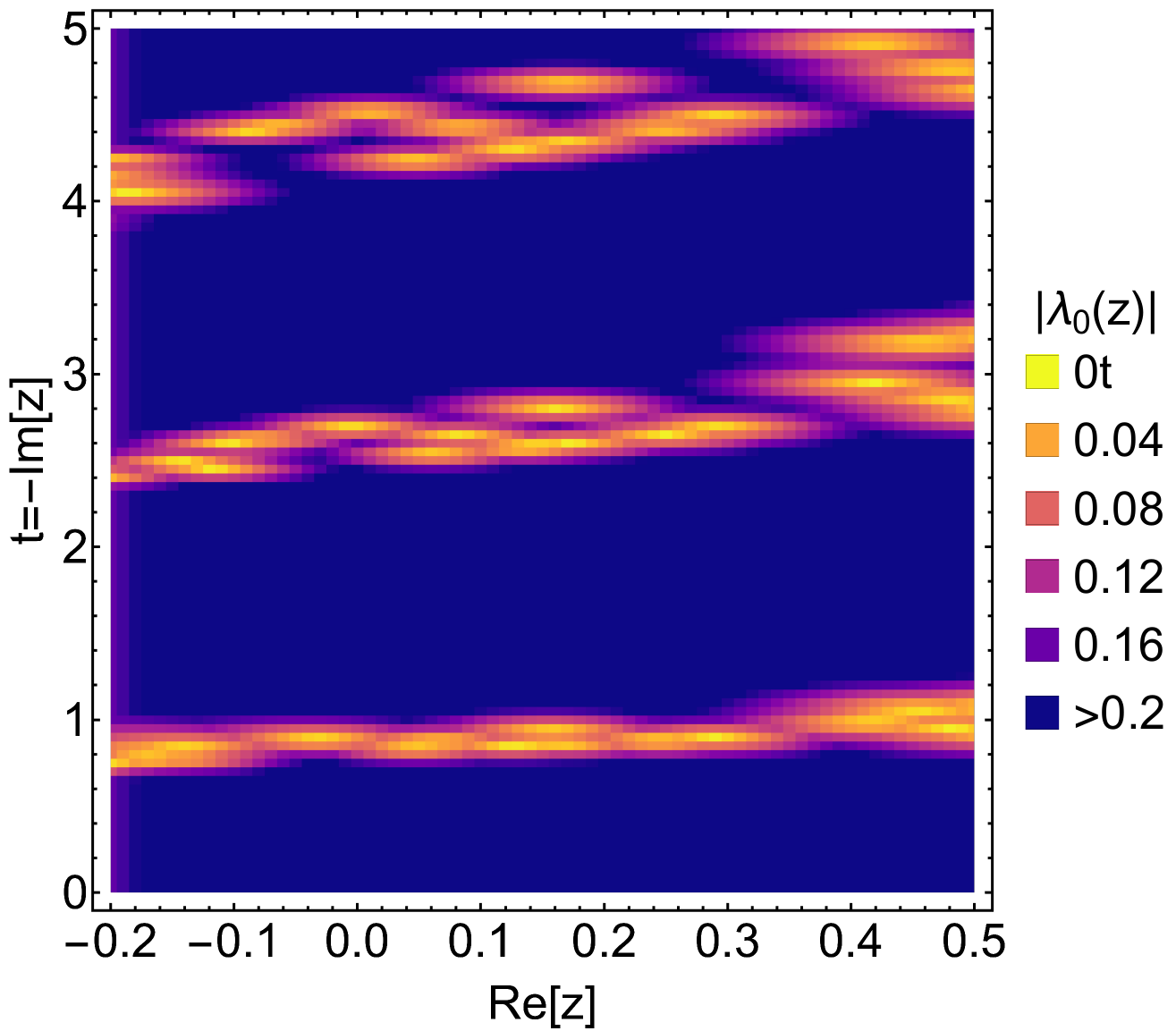}
    \caption{Here we show a proxy for the Fisher zeroes in the complex time plane at a system size $N=40^2$ for the more general two-dimensional model, see appendix \ref{app:sq} for details. Plotted is the value of the lowest eigenvalue of $M(z)$, $|\lambda_0(z)|$, for regions where this is close to zero we may expect Fisher zeroes to occur.}
    \label{fig:Sq_Fisher}
\end{figure}

Fig.~\ref{fig:Sq_Fisher} shows a proxy for the Fisher zeroes. Fisher zeroes occur when $\lambda_0(z)=0$ and therefore we plot $|\lambda_0(z)|$ and look for areas in the complex $z$ plane where $|\lambda_0(z)|$ becomes close to zero. As we are working at a finite size of $N=40^2$ we can not see the exact Fisher zeroes, but regions where they are likely to occur can be seen in Fig.~\ref{fig:Sq_Fisher}, crossing also the real time axis. We therefore expect DQPTs to occur.

\begin{figure}
    \centering
    \includegraphics[width=0.9\columnwidth]{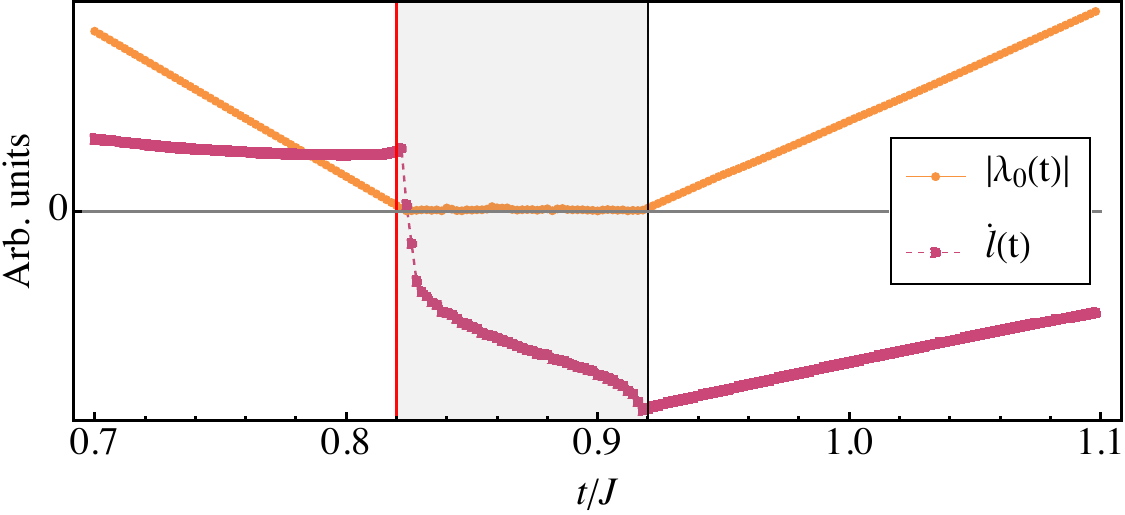}
    \caption{The derivative of the return rate $\dot{l}(t)$ for our exemplary quench in the four-band two-dimensional model $\nu:0\to 1$, see the main text for details. The extent of the critical region, shaded grey, is marked by the region in which the smallest eigenvalue $|\lambda_0(t)|$ is close to zero as we can no longer find the critical region by solving an equivalent of Eq.~\eqref{zns}. The eigenvalues and $\dot{l}(t)$ are calculated for a system size of $N=1600^2$. The cusps in $\dot{l}(t)$ at the critical times are clearly visible.}
    \label{fig:Sq_DQPTd}
\end{figure}

For a complicated model such as the one considered here we can not calculate analytically the critical times, but from $|\lambda_0(t)|$ we can again find a good proxy. Critical times should occur when $|\lambda_0(t)|$ becomes close to zero, and leaves zero. In between we will have a critical region where $|\lambda_0(t)|\approx 0$, if there are any additional critical times inside this region this method will unfortunately not demonstrate them. From this method we can find a reasonable approximation for the critical times. In Fig.~\ref{fig:Sq_DQPTd} we plot $|\lambda_0(t)|$ and the derivative of the return rate $\dot l(t)$. As expected $\dot l(t)$ shows cusps at the edges of the critical region in line with the prediction of Sec.~\ref{sec:gen}.


%

\end{document}